\documentclass[11pt,a4paper,oneside]{article}
\usepackage{lineno,hyperref}

\usepackage{graphicx}              
\usepackage{epsfig}
\usepackage{amsmath}               
\usepackage{amssymb}
\usepackage{epstopdf}
\usepackage{verbatim}
\usepackage{mathptmx}
\usepackage{mathalpha}
\usepackage{mathtools}
\usepackage[scale=0.80]{geometry}
\usepackage{graphicx, times}
\usepackage{amsfonts}              
\usepackage{amsthm}                
\usepackage{multicol}
\usepackage{algorithm}
\usepackage{algorithmic}
\usepackage{varwidth}
\usepackage{parskip}
\usepackage{hyperref}
\usepackage{rotating}
\usepackage[numbers,sort&compress]{natbib}
\usepackage{multirow}
\usepackage{pdflscape}
\usepackage[numbers]{natbib}
\usepackage{caption}
\usepackage{xcolor}
\usepackage{color}
\usepackage{comment}
\usepackage{subcaption}
\newcommand*{\rom}[1]{\expandafter\@\romannumeral #1}

\newcommand{\bea}{\begin{eqnarray}}
	\newcommand{\eea}{\end{eqnarray}}
\newcommand{\bee}{\begin{eqnarray*}}
	\newcommand{\eee}{\end{eqnarray*}}

\begin{document}
\author{Khomesh R. Patle$^{}$\footnote{khomeshpatle5@gmail.com}, G. P. Singh$^{}$\footnote{gpsingh@mth.vnit.ac.in}
\vspace{.3cm}\\
${}^{}$ Department of Mathematics,\\ Visvesvaraya National Institute of Technology, Nagpur, 440010, Maharashtra, India.
\vspace{.3cm}
\date{}}
\title{Reconstructing the cosmic expansion in $f(R, L_{m})$ gravity via parametrized Hubble function constraints}
\maketitle
\begin{abstract} \noindent
We probe the cosmic expansion scenario within the framework of $f(R, L_{m})$ gravity by employing a well-motivated functional form of $f(R, L_{m}) = \frac{R}{2} + L_{m}^{\lambda}$. Specifically, we introduce three novel cosmological models formulated in terms of the redshift-dependent Hubble parameter $H(z)$, offering deeper insights into the underlying cosmic dynamics. The models are further utilized to investigate the expansion history of the universe and the evolution of several cosmological parameters. By using the Bayesian methods based on the $\chi^{2}$-minimization technique, the median values of the model parameters are determined for the cosmic chronometer (CC) and joint (CC+Pantheon) datasets. A comprehensive study of the deceleration parameter, energy density, pressure and the equation of state parameter is carried out to understand the universe's evolution. Additionally, the validity of the energy conditions and the behavior of the statefinder diagnostic are thoroughly examined. Finally, the thermodynamic viability of the models is confirmed through the evolution of temperature and entropy density, and the estimated age of the universe further exemplifies good agreement with late-time astronomical observations.
\end{abstract}

\section{Introduction}\label{sec:1}
Astronomical observations~\cite{1998AJ....116.1009R,1999ApJ...517..565P,2020A&A...641A...6P} have confirmed that the universe is undergoing an accelerated expansion, revealing one of the most intriguing aspects of the observable cosmos. This discovery has sparked significant interest among theoretical cosmologists to construct cosmological scenarios capable of addressing this accelerated phase of cosmic evolution. In response, researchers have pursued two primary approaches: modifying Einstein’s field equations or exploring alternative theories of gravitation to gain deeper insight into the underlying mechanisms driving this phenomenon. Despite extensive efforts, the true nature of the unidentified energy component responsible for this acceleration remains elusive and is commonly known as dark energy. To account for the observed acceleration, various cosmological models have been proposed, many of which are based on dark energy components possessing negative pressure and positive energy density. Among these, the cosmological constant $\Lambda$ stands out as one of the most prominent candidates~\cite{weinberg1989cosmological}. However, despite its remarkable success in describing the observed accelerated expansion of the universe, the $\Lambda$CDM model suffers from serious theoretical issues, particularly the fine-tuning and coincidence problems~\cite{weinberg1989cosmological,di2021realm,carroll2001cosmological}. In recent decades, considerable attention has been directed toward addressing these challenges through alternative theoretical frameworks. One widely adopted strategy involves the study of modified gravity theories. In this context, $f(R)$ gravity~\cite{Kerner,buchdahl1970non} has emerged as an important modification of General Relativity, offering a broader theoretical foundation to better understand cosmic dynamics and the formation of large-scale structures. Along with these developments, several modified gravitational frameworks have been extensively investigated~\cite{nojiri2011unified,nojiri2017modified,harko2011f,cai2016f,capozziello2011cosmography,capozziello2019extended,bamba2010finite,capozziello2023role,lalke2023late,kotambkar2017anisotropic,singh2002viscous,goswami2024flrw,patle2026dynamical,varela2025cosmological,singh2024conservative,singh2022cosmic,hulke2020variable,singh505abc,shukla2025multi}, highlighting the sustained effort to resolve fundamental problems associated with the standard cosmological paradigm.
\vspace{0.2cm}\\
A theoretical framework has been proposed by introducing a coupling between the Ricci scalar $R$ and the matter Lagrangian density $L_m$, leading to the formulation of $f(R, L_m)$ gravity, which incorporates a non-minimal interaction between geometry and matter. As emphasized by Harko and Lobo~\cite{harko2010f}, this approach provides a more general and flexible setting for exploring the intricate relationship between matter and spacetime geometry, thereby opening new avenues for understanding their mutual interplay. However, such an approach is inconsistent with the equivalence principle, while solar system observations place strong constraints on these models~\cite{Faraoni2004pi,zhang2007behavior,bertolami2008general,Rahaman2009solar}. Nevertheless, the $f(R, L_m)$ framework remains a mathematically consistent and physically motivated extension of gravitational theory, formulated within the context of Riemannian geometry~\cite{nojiri2004gravity,allemandi2005dark,manna2023gravity}. It provides valuable insights into the coupling between matter and geometry in both cosmological and astrophysical contexts. A key distinction between $f(R, L_m)$ gravity and the standard $f(R)$ theory arises in the presence of matter, where the field equations differ significantly due to the explicit matter-geometry coupling. However, in the vacuum limit, where matter contributions vanish, both theories reduce to equivalent mathematical forms~\cite{harko2010f,lobo2015extended}. A substantial body of work has been devoted to investigating the cosmological viability of $f(R, L_{m})$ gravity from multiple perspectives. In particular, various studies have explored its implications for late-time cosmic acceleration, observationally constrained dark energy models, bouncing cosmologies and exotic astrophysical structures such as wormholes~\cite{jaybhaye2022cosmology,doi:10.1142/S0219887823501050,koussour2024bouncing,mustafa2024dynamical}. More broadly, extensive investigations have addressed its theoretical structure and observational consequences across diverse cosmological settings~\cite{myrzakulova2024investigating,kavya2022constraining,devi2024constraining,zeyauddin2024anisotropic,maurya2024bianchi,sahlu2024cosmology,bhardwaj2025cosmological,jaybhaye2022constraints,shukla2025dynamical,myrzakulov2024linear,singh2024consequence,chaudhary2024existence,wu2014constraints,kavya2023static}, significantly advancing our understanding of matter-geometry coupling in gravitational physics.
\vspace{0.2cm}\\
The Hubble parameter $H$ is a fundamental quantity that characterizes the expansion rate of the universe and provides direct insight into its dynamical evolution. Being directly constrained by observational probes such as cosmic chronometers and Type Ia supernovae, parametrizations of $H$ offer a robust and model-independent framework to investigate cosmic expansion~\cite{Shafieloo,Amendola}. In the literature, several functional forms of the Hubble parameter have been proposed to describe the transition from an early decelerated phase to the present accelerated epoch; however, many of these parametrizations are either overly restrictive or lack the flexibility required to capture the full expansion history across different cosmic times~\cite{Koussour,RitikaNagpal,he2024new,RoyGoswami}. In this context, a parametrization of the Hubble parameter $H(t)$ is employed and a systematic analysis is performed by introducing three new and previously unexplored cases, each corresponding to a distinct cosmological model. The explicit form of the parametrization is presented in the subsequent section. This framework successfully reproduces a wide range of expansion behaviors and naturally accommodates smooth transitions between various stages of the universe’s evolution, offering a flexible framework for studying the expansion history of the universe in light of observational data.
\vspace{0.2cm}\\
The framework of $f(R, L_m)$ gravity, characterized by a non-minimal interaction between matter and geometry, provides a useful approach for investigating the late-time dynamics of the universe. In this work, we consider a specific functional form given by $f(R, L_m)=\frac{R}{2}+L_m^{\lambda}$, where $\lambda$ is a free model parameter. Within this setup, the adopted Hubble parameter parametrization leads to three distinct cosmological models, each corresponding to a different cosmological evolution. The parameter estimation is carried out using the cosmic chronometer (CC) dataset along with the joint (CC+Pantheon) observational data. The main aim of this study is to analyze the late-time accelerated expansion of the universe through the evolution of key cosmological parameters and to examine the viability of the proposed models.
\vspace{0.2cm}\\
The present work is structured into seven sections as follows: Section~(\ref{sec:2}) briefly outlines the mathematical foundations of the $f(R, L_{m})$ gravity framework. The field equations corresponding to the Friedmann-Lemaître-Robertson-Walker (FLRW) metric are presented in Section~(\ref{sec:3}). In Section~(\ref{sec:4}), the cosmological expansion rate is studied through parametric representations of the Hubble parameter $H(z)$, leading to the construction of three distinct cosmological models. Section~(\ref{sec:5}) is devoted to the observational analysis, where constraints on the model parameters are obtained using Bayesian statistical methods based on cosmic chronometer (CC) data and a combined dataset comprising CC and supernovae Type Ia observations. In Section~(\ref{sec:6}), we study the physical and dynamical features of the proposed models, focusing on the evolution of key quantities such as the deceleration parameter, energy density, pressure, and the equation of state (EoS) parameter. The energy conditions and statefinder diagnostic are further employed to assess the physical viability of the models and to distinguish their evolution from standard cosmological scenarios. Moreover, the models are tested for thermodynamic consistency by analyzing the evolution of temperature and entropy density, together with the estimation of the universe’s age. To conclude, Section~(\ref{sec:7}) summarizes the main findings of the work and presents the concluding remarks.
\section{Mathematical foundations of $f(R, L_{m})$ gravity theory}\label{sec:2}
Here, we present the basic formulation of $f(R, L_m)$ gravity with the corresponding action defined as~\cite{harko2010f}:
\begin{equation}{\label{1}}
	S=\int f(R, L_{m})\sqrt{-g} d^{4}x,
\end{equation}
where $L_m$ represents the matter Lagrangian density and $R$ denotes the Ricci scalar. We can define Ricci scalar ($R$) by using the Ricci tensor ($R_{\mu\nu}$) and the metric tensor ($g^{\mu\nu}$) as follows
\begin{equation}{\label{2}}
	R=g^{\mu \nu}R_{\mu \nu}, 
\end{equation}
where the $R_{\mu \nu}$ is
\begin{equation}{\label{3}}
	R_{\mu \nu}= \partial_{c} \Gamma^{c}_{\mu \nu}-\partial_{\mu} \Gamma^{c}_{c\nu}+\Gamma^{c}_{\mu \nu}\Gamma^{d}_{dc}-\Gamma^{c}_{\nu d} \Gamma^{d}_{\mu c},
\end{equation}
and $ \Gamma^{\alpha}_{\beta \gamma} $ signifies the Levi-Civita connection expressed as
\begin{equation}{\label{4}}
	\Gamma^{\alpha}_{\beta \gamma}= \frac{1}{2}g^{\alpha c}\left(\frac{\partial g_{\gamma c}}{\partial x^{\beta}}+\frac{\partial g_{c \beta}}{\partial x^{\gamma}}-\frac{\partial g_{\beta \gamma }}{\partial x^{c}} \right).
\end{equation}
By varying the action (\ref{1}) with respect to the metric tensor $g_{\mu\nu}$, the corresponding field equations can be expressed as,
\begin{equation}{\label{5}}
	\frac{\partial F}{\partial R}R_{\mu \nu}+(g_{\mu \nu} \square -\nabla_{\mu}\nabla_{\nu})\frac{\partial F }{\partial R}-\frac{1}{2}\left( F-\frac{\partial F}{\partial L_{m}}L_{m}\right)g_{\mu \nu}=\frac{1}{2}\left(\frac{\partial F}{\partial L_{m}}\right)T_{\mu \nu}, 
\end{equation}
where $\square=g^{\mu \nu}\nabla_{\mu}\nabla_{\nu}$, $ F=f(R, L_{m})$ and $T_{\mu \nu}$ is defined as the energy-momentum tensor for perfect fluid
as
\begin{equation}{\label{6}}
	T_{\mu \nu}=\frac{-2}{\sqrt{-g}}\frac{\delta(\sqrt{-g}L_{m})}{\delta g^{\mu \nu}}.
\end{equation}
From the above field equations, a connection between the Ricci scalar $(R)$, the trace of the energy-momentum tensor $(T)$ and the matter Lagrangian density $(L_{m})$ can be derived as
\begin{equation}{\label{7}}
	R\left(\frac{\partial F}{\partial R}\right)  +2\left(\frac{\partial F}{\partial L_{m}}L_{m}-F\right)+ 3\square \frac{\partial F}{\partial R}=\frac{1}{2}\left(\frac{\partial F}{\partial L_{m}}\right)T,
\end{equation}
where $ \square I=\frac{1}{\sqrt{-g}}\partial_{\mu}(\sqrt{-g}g^{\mu \nu} \partial_{\nu}I)$ defines the d'Alembertian operator acting on any arbitrary function $I$. To investigate Eq. (\ref{5}) further, the covariant derivative may be reformulated in terms of the energy-momentum tensor as follows:
\begin{equation}{\label{8}}
	\nabla^{\mu}T_{\mu \nu}=2\nabla^{\mu} \log\left(\frac{\partial F}{\partial L_{m}}\right) \frac{\partial L_{m}}{\partial g^{\mu \nu}}.
\end{equation} 
In the subsequent section, the equations of motion are obtained for the FLRW spacetime.
\section{Equations governing $f(R, L_{m})$ gravity}\label{sec:3}
On large cosmological scales, the observable universe is well approximated as homogeneous and isotropic. To model this feature, we consider the flat Friedmann-Lemaître-Robertson-Walker (FLRW) metric~\cite{partridge2004introduction}, which is formulated as
\begin{equation}{\label{9}}  
	ds^{2}=-dt^{2}+a^{2}(t) \left( dx^{2}+ dy^{2}+ dz^{2}\right),
\end{equation}
where the cosmic expansion scale factor $a(t)$ is defined as a function of cosmic time ($t$). Based on the line element (\ref{9}), the non-vanishing
components of the Christoffel symbols are:
\begin{equation}{\label{10}}
	\Gamma^{0}_{pq}= -\frac{1}{2}g^{00} \  \frac{\partial g_{pq}}{\partial x^{0}}, \  \  \ \ \Gamma^{r}_{0q}=\Gamma^{r}_{q0}= \frac{1}{2}g^{r\lambda} \  \frac{\partial g_{q \lambda}}{\partial x^{0}},
\end{equation}
where $p, \ q, \ r = 1, \ 2, \ 3$ denote the spatial indices. From Eq. (\ref{3}), the non-vanishing Ricci tensor components and the corresponding Ricci scalar are obtained as follows: 
\begin{equation*}
	R^{0}_{0}=3\frac{\ddot{a}}{a}, \  \ R^{1}_{1}=R^{2}_{2}=R^{3}_{3}=\frac{\ddot{a}}{a}+2\left(\frac{\dot{a}}{a}\right)^{2},            
\end{equation*}
\begin{equation}{\label{11}}
	R=6 \left[ \left(\frac{\dot{a}}{a}\right)^{2}+ \left( \frac{\ddot{a}}{a}\right) \right] = 6 \left(\dot{H}+ 2H^{2} \right),           
\end{equation}
where $H = \frac{\dot{a}}{a}$ denotes the Hubble parameter describing the expansion rate of the universe. For a perfect fluid, the energy-momentum tensor is taken as
\begin{equation}{\label{12}}
	T_{\mu \nu} = (\mathit{p} + \rho) u_{\mu} u_{\nu} + p g_{\mu\nu},
\end{equation}
where $p$ and $\rho$ represent the isotropic pressure and energy density of the cosmic fluid, respectively and the four-velocity components are taken as $u^\mu = (1, 0, 0, 0)$, which obey the normalization relation $u^\mu u_\mu = -1$. The Friedmann equations governing the dynamical evolution of the universe in $f(R, L_m)$ gravity can be written as follows:
\begin{equation}{\label{13}}
	\frac{1}{2} (F -F_{L_m} L_m - F_R R) + 3H  \dot{F}_R +3H^2 F_R  = \frac{1}{2} F_{L_m} \rho
\end{equation}
and 
\begin{equation}{\label{14}}
	3H^2 F_R +\dot{H} F_R- \ddot{F}_R - 3H \dot{F}_R + \frac{1}{2} (F_{L_m} L_m -F) = \frac{1}{2} F_{L_m} p.
\end{equation}
The resulting field equations provide the foundation for determining the cosmic expansion rate, which governs the dynamical evolution and transition history of the universe.
\section{Cosmological expansion rate and background dynamics in $f(R, L_{m})$ model} \label{sec:4}
Here, we consider a specific functional form of the $f(R, L_m)$ gravity model~\cite{harko2014generalized}, given by:
\begin{equation}{\label{15}}
	f(R, L_m) = \frac{R}{2} + L_m^{\lambda},
\end{equation}
where $\lambda$ is an arbitrary constant. For the specified functional form of the $f(R, L_m)$ model with $L_m = \rho$~\cite{harko2015gravitational}, combining Eqs.~(\ref{13}) and (\ref{15}) yields
\begin{equation} {\label{16}}
	3H^2 = (2\lambda - 1)\rho^{\lambda},
\end{equation}
furthermore, by using $R = 6\left(\dot{H} + 2H^{2}\right)$, Eqs.~(\ref{14}) and (\ref{15}) give
\begin{equation} {\label{17}}
	2\dot{H} + 3H^2 = (\lambda - 1)\rho^{\lambda} - \lambda p \rho^{\lambda - 1}.
\end{equation}
Mathematically, the Equation of State (EoS) parameter is defined as $\omega = \frac{p}{\rho}$. Utilizing Eqs.~(\ref{16}) and (\ref{17}) together with the relation $\dot{H} = -H(1+z)\frac{dH}{dz}$, we derive the EoS parameter ($\omega$) in the form:
\begin{equation}{\label{18}}
	\omega = \frac{2(2\lambda - 1)(1+z)H' - 3\lambda H}{3\lambda H}.
\end{equation}
The above expression of the EoS parameter $\omega$ in terms of the Hubble parameter $H(z)$ provides a useful framework to study the expansion dynamics of the universe. This relation enables us to analyze and reconstruct the evolution of the cosmic equation of state for different cosmological scenarios through suitable choices of $H(z)$. To proceed further, we introduce particular parametrized forms of the Hubble parameter $H(z)$ in the following subsection.
\subsection{Analytical forms of the Hubble parameter $H(z)$}\label{sec:4.1}
The Hubble parameter is an essential quantity for understanding the expansion behavior of the universe and is widely employed in cosmological studies developed under different theories of gravitation. Besides theory-dependent descriptions, the cosmic expansion history can also be investigated using model-independent techniques that avoid assuming a particular gravitational framework~\cite{shafieloo2013model}. In these approaches, suitable parametrizations are introduced for cosmological quantities such as the Hubble parameter, pressure and energy density in order to analyze the dynamical evolution of the universe. Such parametrized descriptions are particularly useful for examining the transition from the early decelerated phase of expansion to the presently accelerating epoch through quantities including the Hubble parameter, deceleration parameter and EoS parameter. The resulting models may then be constrained using observational datasets. Parametrization-based cosmological studies have attracted considerable attention in the literature due to their usefulness in addressing several open problems, including the initial singularity, horizon problem, persistent deceleration scenario and the Hubble tension~\cite{banerjee2005acceleration,cunha2008transition,escamilla2022dynamical}. In this direction, the observed late-time acceleration of the universe may be described through an appropriate functional choice of the Hubble parameter $H(z)$~\cite{RoyGoswami,myrzakulov2023quintessence,NagpalPacif,yadav2024reconstructing}. Motivated by this perspective, we employ a parametrized form of the Hubble parameter proposed in Ref.~\cite{pacif2017reconstruction}, where $H$ is expressed explicitly in terms of cosmic time $t$ as follows:
\begin{equation}{\label{19}}
	H(t)= \frac{\kappa_{2} t^{n}}{(t^{m}+\kappa_{1})^{\tau}},
\end{equation}
where $\kappa_{1}$, $\kappa_{2}$ $\neq 0$ and $m$, $n$, $\tau$ are real model parameters. The constants $\kappa_{1}$ and $\kappa_{2}$ have the dimensions of time. Particular choices of the parameters $m$, $n$ and $\tau$ lead to distinct cosmological models. In this work, we introduce three new cosmological models corresponding to $(m=3, \tau=1, n=-1)$, $(m=4, \tau=1, n=-1)$ and $(m=5, \tau=1, n=-1)$, derived from the functional form of the Hubble parameter given in Eq.~(\ref{19}). These models are constructed by selecting specific integral and non-integral values of the parameters $m$, $n$, and $\tau$. Notably, these choices exhibit the capability to describe the cosmological phase transition for negative values of $\kappa_{1}$ and $\kappa_{2}$. The resulting models are summarized in Table~(\ref{table:1}). Here, $\eta$ is an integration constant that also plays a significant role in the cosmic evolution. In the present analysis, we focus on these three models, namely Model-I, Model-II and Model-III, and investigate their cosmological behavior in detail.
\begin{table}[h!]
	\centering
	\caption{The models}
	\label{table:1}
    \renewcommand{\arraystretch}{1.0}  
    \fontsize{16pt}{7pt}\selectfont  
		\begin{tabular}{|c|c|c|c|c|}
			\hline
			Models & $H(t)$ & $a(t)$ & $q(t)$ \\
			\hline
			Model-I & $\dfrac{\kappa_2}{t(\kappa_1-t^{3})}$ & $\eta \left(\dfrac{t^{3}}{\kappa_1-t^{3}}\right)^{\frac{\kappa_2}{3\kappa_1}}$ & $-1 + \dfrac{\kappa_1}{\kappa_2} - \dfrac{4}{\kappa_2} t^{3}$ \\
			\hline
			Model-II & $\dfrac{\kappa_2}{t(\kappa_1-t^4)}$ & $\eta \left(\dfrac{t^4}{\kappa_1-t^4}\right)^{\frac{\kappa_2}{4 \kappa_1}}$ & $-1 + \dfrac{\kappa_1}{\kappa_2} - \dfrac{5}{\kappa_2} t^4$ \\
			\hline
			Model-III & $\dfrac{\kappa_2}{t(\kappa_1-t^5)}$ & $\eta \left(\dfrac{t^5}{\kappa_1-t^5}\right)^{\frac{\kappa_2}{5 \kappa_1}}$ & $-1 + \dfrac{\kappa_1}{\kappa_2} - \dfrac{6}{\kappa_2} t^5$ \\
			\hline
	\end{tabular} 
\end{table}
\vspace{.2cm}\\
It can be observed that, for all three models, the Hubble parameter and the scale factor diverge at a finite cosmic time, indicating the occurrence of a Big Rip singularity in the near future. For Model-I, the singularity occurs at $t = t_s = \sqrt[3]{\kappa_{1}}$, while for Model-II and Model-III, it appears at $t = t_s = \sqrt[4]{\kappa_{1}}$ and $t = t_s = \sqrt[5]{\kappa_{1}}$, respectively. Furthermore, the transition from a decelerating to an accelerating phase of the universe occurs at $t_{tr} = \sqrt[3]{\frac{\kappa_{1} - \kappa_{2}}{4}}$ for Model-I, $t_{tr} = \sqrt[4]{\frac{\kappa_{1} - \kappa_{2}}{5}}$ for Model-II, and $t_{tr} = \sqrt[5]{\frac{\kappa_{1} - \kappa_{2}}{6}}$ for Model-III. These results imply that the condition $\kappa_{1} > \kappa_{2}$ must be satisfied for the transition time to be physically meaningful.
\vspace{.2cm}\\
For observational investigations, it is generally more convenient to formulate cosmological quantities in terms of the redshift variable ($z$). However, the parameters employed in the present analysis are expressed as functions of cosmic time ($t$). Hence, establishing the relationship between cosmic time and redshift becomes important. Accordingly, the $t$-$z$ relations corresponding to the three considered models are derived below:
\begin{equation}{\label{20}}
	t(z)= \sqrt[3]{\kappa_{1}} \left[1+(\eta (1+z))^{\frac{3\kappa_{1}}{\kappa_{2}}} \right]^{\frac{-1}{3}}  ~~~\qquad \qquad \qquad \qquad     \text{(for Model-I)}
\end{equation}
\begin{equation}{\label{21}}
	t(z)= \sqrt[4]{\kappa_{1}} \left[1+(\eta (1+z))^{\frac{4\kappa_{1}}{\kappa_{2}}} \right]^{\frac{-1}{4}}  ~~~\qquad \qquad \qquad \qquad     \text{(for Model-II)}
\end{equation}
\begin{equation}{\label{22}}
	t(z)= \sqrt[5]{\kappa_{1}} \left[1+(\eta (1+z))^{\frac{5\kappa_{1}}{\kappa_{2}}} \right]^{\frac{-1}{5}}  ~~~\qquad \qquad \qquad \qquad     \text{(for Model-III)}
\end{equation}
Equations~(\ref{20}) to (\ref{22}) are characterized by the three parameters $\eta$, $\kappa_{1}$ and $\kappa_{2}$. In order to make the formulation more compact and reduce the number of independent parameters, we introduce the dimensionless quantity $\zeta=\kappa_{1}/\kappa_{2}$. This redefinition considerably simplifies the mathematical treatment of the models. As a result, the corresponding expressions of the Hubble parameter for all three models can be represented in terms of the redshift parameter $z$ as follows:
\begin{equation}{\label{23}}
	H(z)= H_{0}(1+\eta^{3\zeta})^{\frac{-4}{3}} (1+z)^{-3\zeta} \left[1+(\eta (1+z))^{3\zeta} \right]^{\frac{4}{3}}     \qquad \qquad \qquad \qquad     \text{(for Model-I)}
\end{equation}
\begin{equation}{\label{24}}
H(z)= H_{0}(1+\eta^{4\zeta})^{\frac{-5}{4}} (1+z)^{-4\zeta} \left[1+(\eta (1+z))^{4\zeta} \right]^{\frac{5}{4}}     \qquad \qquad \qquad \qquad     \text{(for Model-II)}
\end{equation}
\begin{equation}{\label{25}}
	H(z)= H_{0}(1+\eta^{5\zeta})^{\frac{-6}{5}} (1+z)^{-5\zeta} \left[1+(\eta (1+z))^{5\zeta} \right]^{\frac{6}{5}}     \qquad \qquad \qquad \qquad     \text{(for Model-III)}
\end{equation}
where $H_{0}$ denotes the Hubble parameter at redshift $z = 0$, while $\zeta$ and $\eta$ are free model parameters to be constrained using observational data.
\section{Observational constraints and results}\label{sec:5}
This section is devoted to the observational assessment of the proposed expansion histories given in Eqs.~(\ref{23}) to (\ref{25}) within the framework of Bayesian statistical analysis. The model parameters $H_{0}$, $\zeta$ and $\eta$, appearing in the parametrized forms of the Hubble parameter for the three models, are constrained using observational datasets, including the cosmic chronometer (CC) dataset and the Type Ia supernova (Pantheon) dataset. We further combine these datasets to form a joint sample, referred to as the CC+Pantheon dataset. In the present analysis, the parameter estimation is performed through $\chi^{2}$ minimization within the Markov chain Monte Carlo (MCMC) technique, implemented using the emcee Python package~\cite{foreman2013emcee}.
\subsection{Cosmic chronometer observational dataset}\label{sec:5.1}
\begin{figure}[ht]
	\centering
	\includegraphics[width=14cm, height=6cm]{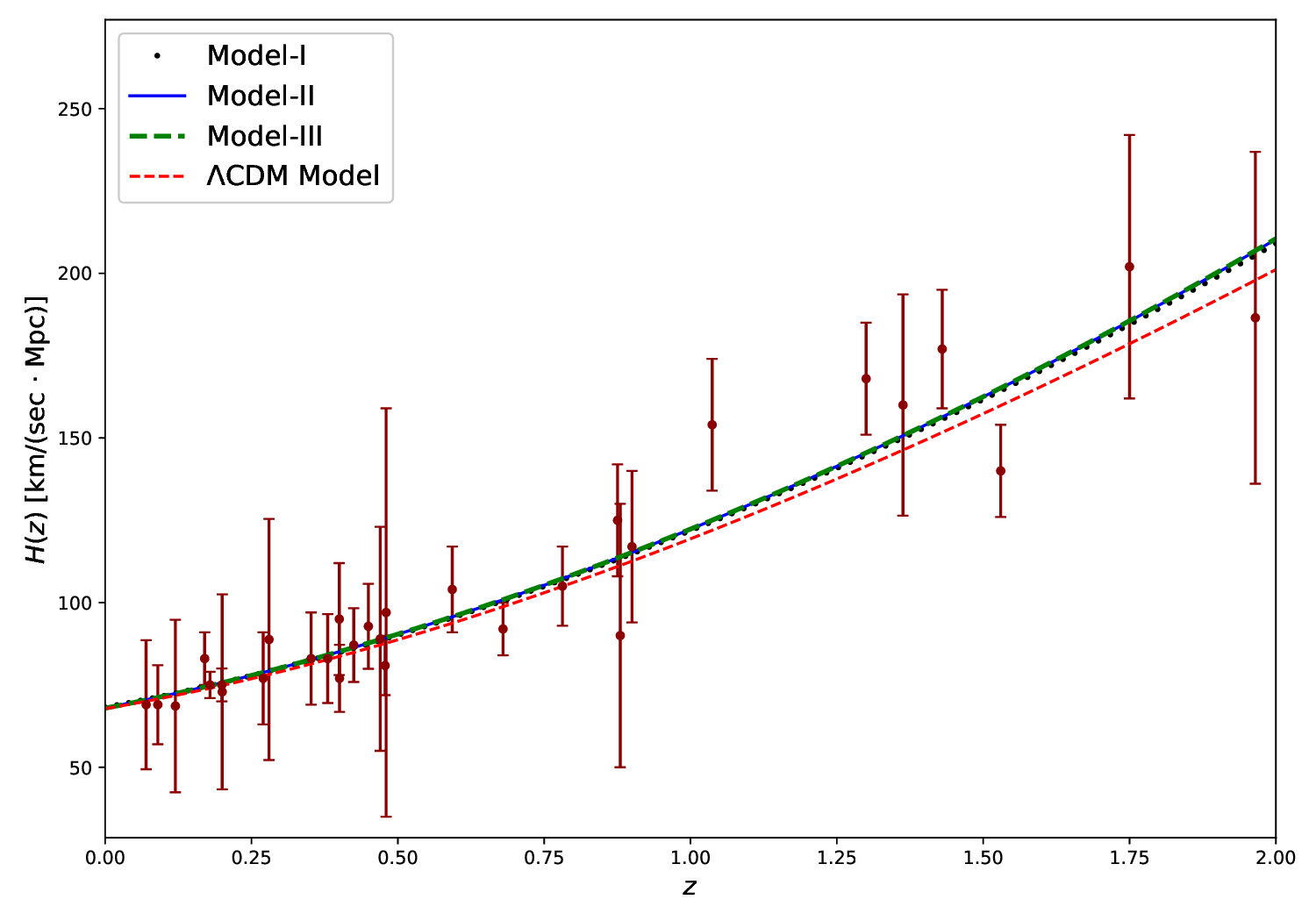}
	\caption{Comparison of the best-fit $H(z)$ profiles of the proposed models with the $\Lambda$CDM model.}
	\label{fig:1}
\end{figure}

To place constraints on the model parameters, we examine the observational consistency of the proposed cosmological framework using cosmic chronometer (CC) data. The analysis is performed with a compilation of 31 CC measurements~\cite{simon2005constraints, sharov2018predictions}, derived through the differential age (DA) method for galaxies spanning the redshift interval $0.07 \leq z \leq 1.965$~\cite{stern2010cosmic, moresco2015raising}. The main purpose of this analysis is to determine the best-fit or median estimates of the model parameters. According to the method proposed by Jimenez and Loeb~\cite{jimenez2002constraining}, the Hubble parameter can be expressed in terms of redshift and cosmic time through the relation $H(z)= -\frac{1}{1+z}\frac{dz}{dt}$. In the present work, the parameters $H_{0}$, $\zeta$ and $\eta$ are constrained within a standard statistical framework by minimizing the chi-squared function ($\chi^{2}$), which is equivalent to maximizing the likelihood function~\cite{singh505abc,mandal2023cosmic}, given by:
\begin{equation}{\label{26}}
	\chi^{2}_{CC}(\theta)=\sum_{i=1}^{31} \frac{[H_{th}(\theta,z_{i})-H_{obs}(z_{i})]^{2} }{ \sigma^{2}_{H(z_{i})}}.  
\end{equation} 
\vspace{0.1cm}\\
Here, $H_{th}$ represents the theoretically predicted value of the Hubble parameter, while $H_{obs}$ denotes its observed value obtained from the data. The quantity $\sigma_{H}$ corresponds to the standard uncertainty associated with the observational measurements.
\vspace{0.1cm}\\
Figure~(\ref{fig:1}) shows the CC data points with their corresponding error bars along with the best-fit Hubble parameter curve.
\subsection{Pantheon supernova observational dataset}\label{sec:5.2}
\begin{center}
	\begin{figure}
		\includegraphics[width=18.5cm, height=18.5cm]{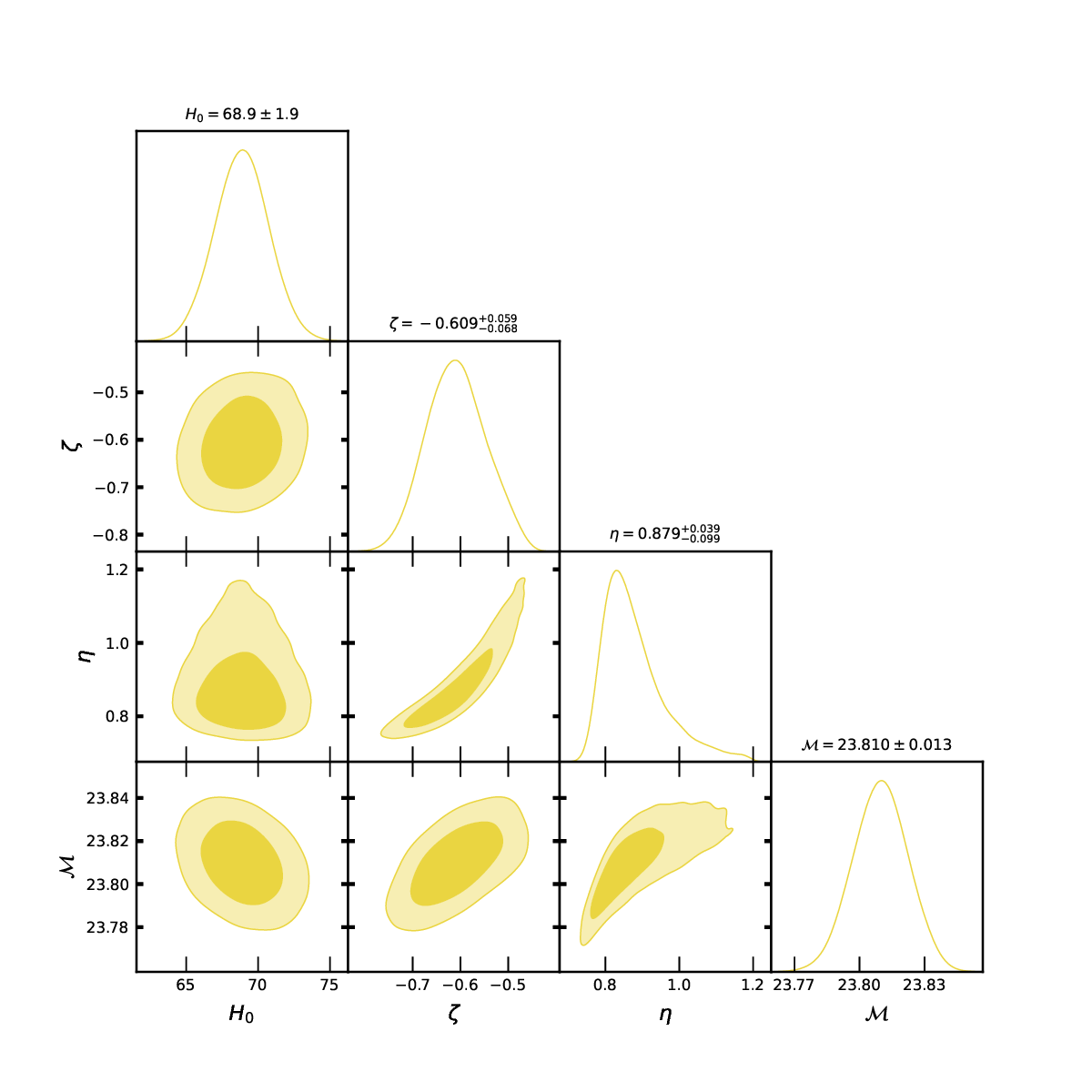}
		\caption{\textbf{For Model-I:} Marginalized $1\sigma$ and $2\sigma$ confidence contours together with the median estimates of $H_{0}$, $\zeta$ and $\eta$ obtained from the joint dataset.}
		\label{fig:2}
	\end{figure}
\end{center}

\begin{center}
	\begin{figure}
		\includegraphics[width=18.5cm, height=18.5cm]{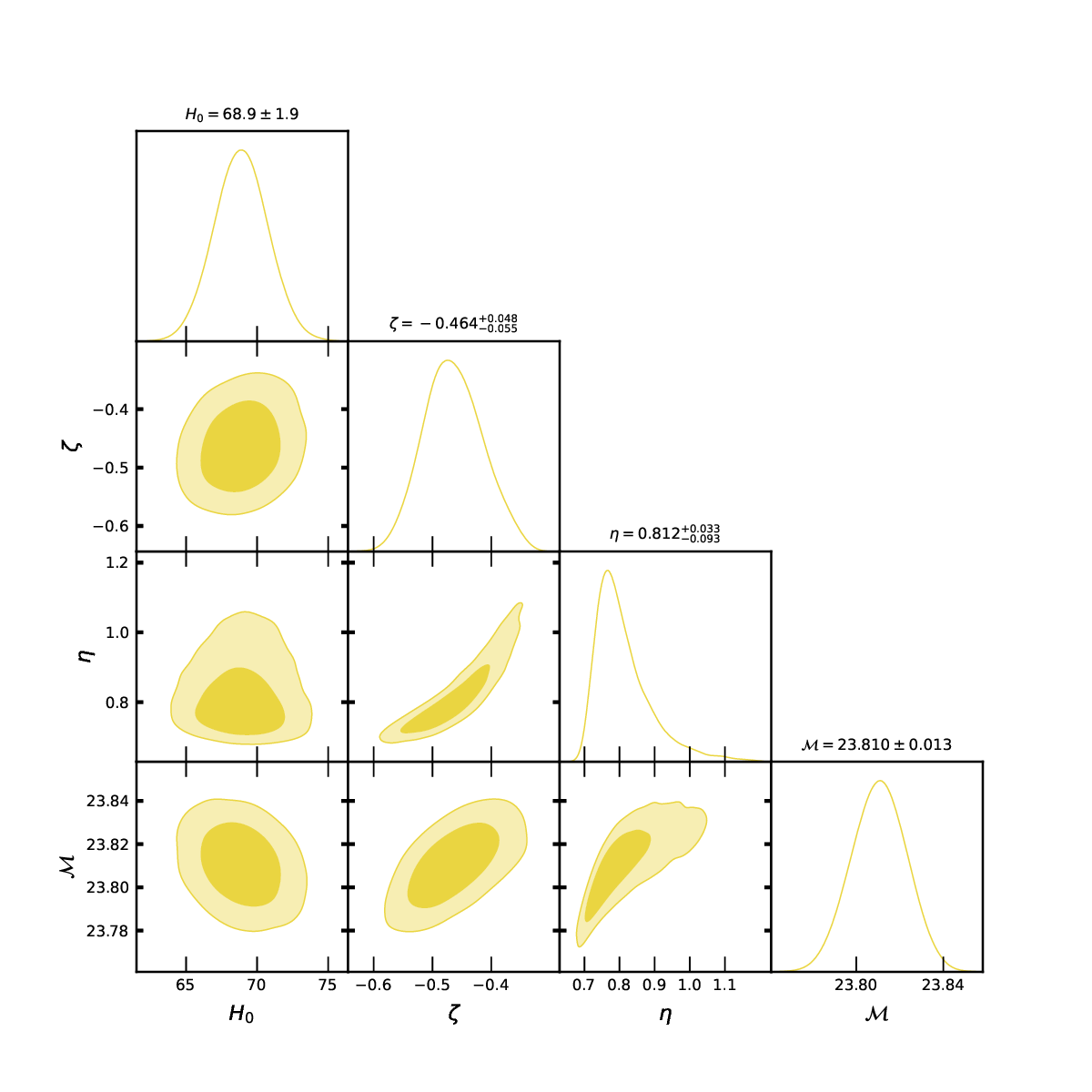}
		\caption{\textbf{For Model-II:} Marginalized $1\sigma$ and $2\sigma$ confidence contours together with the median estimates of $H_{0}$, $\zeta$ and $\eta$ obtained from the joint dataset.}
		\label{fig:3}
	\end{figure}
\end{center}

\begin{center}
	\begin{figure}
		\includegraphics[width=18.5cm, height=18.5cm]{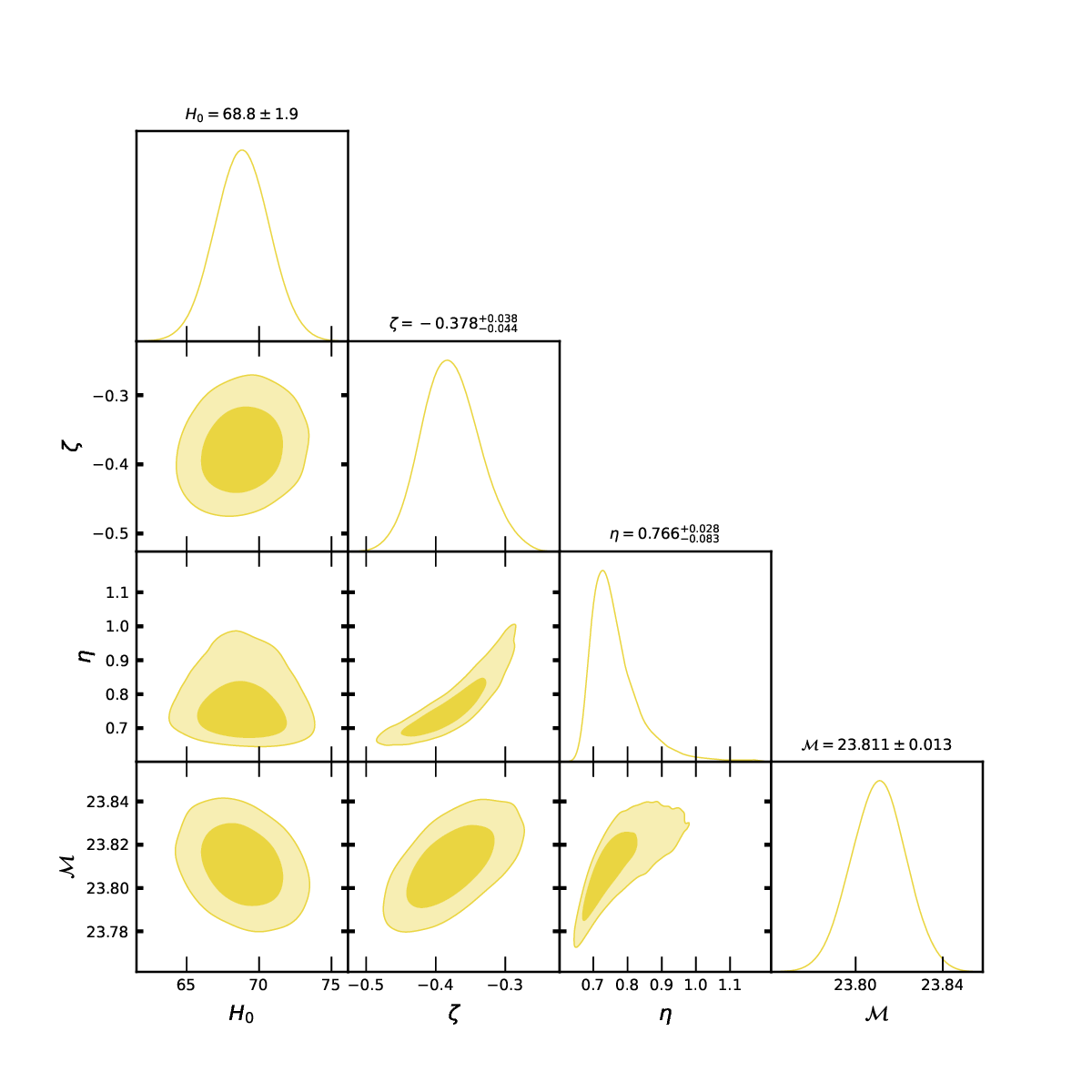}
		\caption{\textbf{For Model-III:} Marginalized $1\sigma$ and $2\sigma$ confidence contours together with the median estimates of $H_{0}$, $\zeta$ and $\eta$ obtained from the joint dataset.}
		\label{fig:4}
	\end{figure}
\end{center}

In the present analysis, we further employ the Pantheon compilation consisting of 1048 Type Ia supernova (SNIa) data points distributed over the redshift interval $0.01 < z < 2.26$~\cite{scolnic2018complete}. The Pantheon sample is constructed by combining observations from several major supernova surveys, such as the CfA1-CfA4 compilations~\cite{riess1999bvri,hicken2009improved}, the Pan-STARRS1 Medium Deep Survey~\cite{scolnic2018complete}, SDSS~\cite{sako2018data}, SNLS~\cite{guy2010supernova} and the Carnegie Supernova Project (CSP)~\cite{contreras2010carnegie}. Within the MCMC framework based on the Pantheon dataset, the theoretical distance modulus $\mu_{th}(z)$ is expressed as follows:
\begin{equation}{\label{27}}
\mu_{th}(z)=M+5log_{10}\left[\frac{d_{L}(z)}{Mpc}\right]+25,
\end{equation}
Here, $M$ corresponds to the absolute magnitude parameter. Furthermore, the luminosity distance $d_{L}(z)$, which carries units of length, is expressed as~\cite{odintsov2018cosmological}
\begin{equation}{\label{28}}
	d_{L}(z)=c(1+z)\int_{0}^{z}\frac{dz'}{H(z')}.
\end{equation}
In this formulation, $z$ indicates the redshift of the Type Ia supernovae (SNIa) evaluated in the cosmic microwave background (CMB) rest frame and $c$ denotes the velocity of light in vacuum. For convenience, the luminosity distance $d_{L}(z)$ is generally rewritten through the dimensionless Hubble-independent quantity $D_{L}(z)=H_{0}d_{L}(z)/c$. Consequently, Eq.~(\ref{27}) takes the form:
\begin{equation}{\label{29}}
	\mu_{th}(z)=M+5log_{10}\left[D_{L}(z)\right]+5log_{10}\left[\frac{c/H_{0}}{Mpc}\right]+25. 
\end{equation}
In the context of the $\Lambda$CDM model, the parameters $M$ and $H_{0}$ are known to exhibit a degeneracy~\cite{ellis2012relativistic,asvesta2022observational}. To incorporate this feature into the analysis, a new parameter $\mathcal{M}$ is introduced by combining $M$ and $H_{0}$, which can be expressed as:
\begin{equation}{\label{30}}
	\mathcal{M}\equiv 25+5log_{10} \left[\frac{c/H_{0}}{Mpc}\right]+M=M+42.38-5log_{10}(h), 
\end{equation}
where the Hubble constant is parameterized as $H_{0}=100 \times h$ $[\text{km}/(\text{sec}.\text{Mpc})]$. Using these quantities, the MCMC analysis is carried out together with the corresponding chi-squared statistic $\chi^{2}$ for the Pantheon compilation, following the formulation presented in~\cite{mandal2023cosmic,asvesta2022observational}:
\begin{equation}{\label{31}}
	\chi^{2}_{P}= \nabla \mu_{i}C^{-1}_{ij}\nabla \mu_{j},
\end{equation}
whereas $\nabla \mu_{i} = \mu_{obs}(z_{i}) - \mu_{th}(z_{i})$, $C_{ij}^{-1}$ corresponds to the inverse covariance matrix, while the expression for $\mu_{th}$ is provided in Eq.~(\ref{29}).
\vspace{0.1cm}\\
The luminosity distance is intrinsically related to the Hubble expansion rate and therefore provides an important observational tool for constraining cosmological parameters. In the present analysis, we utilize the emcee MCMC sampler~\cite{foreman2013emcee} together with the corresponding cosmological relations to carry out a maximum likelihood estimation (MLE) based on the combined CC and Pantheon datasets. For the joint observational study, the total $\chi^{2}$ statistic is constructed as $\chi^{2}_{CC} + \chi^{2}_{P}$. The outcomes of the MCMC fitting procedure are presented in Figures~(\ref{fig:2}) to (\ref{fig:4}), which show the marginalized $1\sigma$ and $2\sigma$ confidence contours as well as the corresponding $1$D posterior distributions, are presented for the combined CC+Pantheon analysis. The inferred median values of the model parameters are summarized in Tables~(\ref{table:2}) to (\ref{table:4}).
\begin{table}[htbp]
	\centering
	\renewcommand{\arraystretch}{3.5}  
	\fontsize{9pt}{9pt}\selectfont  
	\begin{tabular}{|c|c|c|c|c|c|c|c|c|c|}
		\hline
		Dataset & $H_{0}$[Km/(sec.Mpc)] & $\zeta$ & $\eta$ & $\mathcal{M}$ & $q_{0}$ & $z_{t}$ & $\omega_{0}$ & $t_{0}$(Gyr) \\
		\hline
		CC & $68.235^{+0.830}_{-0.835}$ & $-0.608^{+0.028}_{-0.027}$ & $0.882^{+0.066}_{-0.051}$ & - & $-0.5323$ & $0.6547$ & $-0.7015$ & $13.20$ \\
		\hline
		CC+Pantheon  & $68.9^{+1.9}_{-1.9}$  & $-0.609^{+0.059}_{-0.068}$ &  $0.879^{+0.039}_{-0.099}$ &  $23.810^{+0.013}_{-0.013}$ & $-0.5338$ & $0.6523$ & $-0.7022$ & $13.08$ \\
		\hline
	\end{tabular} 
	\caption{ \textbf{For Model-I:} Constraints on the model parameters from the CC and joint datasets, including the current values of $q_{0}$, $\omega_{0}$ and $t_{0}$.}
	\label{table:2}
\end{table}
\begin{table}[htbp]
	\centering
	\renewcommand{\arraystretch}{3.5}  
	\fontsize{9pt}{9pt}\selectfont   
	\begin{tabular}{|c|c|c|c|c|c|c|c|c|c|}
			\hline
			Dataset & $H_{0}$[Km/(sec.Mpc)] & $\zeta$ & $\eta$ & $\mathcal{M}$ & $q_{0}$ & $z_{t}$ & $\omega_{0}$ & $t_{0}$(Gyr) \\
			\hline
			CC & $68.045^{+0.818}_{-0.807}$ & $-0.459^{+0.022}_{-0.022}$ & $0.826^{+0.063}_{-0.049}$ & - & $-0.5108$ & $0.6442$ & $-0.6875$ & $13.20$ \\
			\hline
			CC+Pantheon  & $68.9^{+1.9}_{-1.9}$  & $-0.464^{+0.048}_{-0.055}$ &  $0.812^{+0.033}_{-0.093}$ &  $23.810^{+0.013}_{-0.013}$ & $-0.5254$ & $0.6495$ & $-0.6968$ & $13.05$ \\
			\hline
	\end{tabular} 
     \caption{ \textbf{For Model-II:} Constraints on the model parameters from the CC and joint datasets, including the current values of $q_{0}$, $\omega_{0}$ and $t_{0}$.}
	\label{table:3}
\end{table}
\begin{table}[htbp]
	\centering
	\renewcommand{\arraystretch}{3.5}  
	\fontsize{9pt}{9pt}\selectfont   
	\begin{tabular}{|c|c|c|c|c|c|c|c|c|c|}
		\hline
		Dataset & $H_{0}$[Km/(sec.Mpc)] & $\zeta$ & $\eta$ & $\mathcal{M}$ & $q_{0}$ & $z_{t}$ & $\omega_{0}$ & $t_{0}$(Gyr) \\
		\hline
		CC & $67.942^{+0.802}_{-0.791}$ & $-0.370^{+0.018}_{-0.018}$ & $0.789^{+0.060}_{-0.046}$ & - & $-0.4995$ & $0.6486$ & $-0.6802$ & $13.18$ \\
		\hline
		CC+Pantheon  & $68.8^{+1.9}_{-1.9}$  & $-0.378^{+0.038}_{-0.044}$ &  $0.766^{+0.028}_{-0.083}$ &  $23.811^{+0.013}_{-0.013}$ & $-0.5238$ & $0.6504$ & $-0.6957$ & $13.02$ \\
		\hline
	\end{tabular} 
	\caption{ \textbf{For Model-III:} Constraints on the model parameters from the CC and joint datasets, including the current values of $q_{0}$, $\omega_{0}$ and $t_{0}$.}
	\label{table:4}
\end{table}
\section{Physical aspects and cosmological dynamics of the models}\label{sec:6}
\subsection{Analysis of deceleration parameter}\label{subsec:6.1}
In cosmological studies, the deceleration parameter $q$ is widely used to investigate the dynamical expansion history of the universe. Its value determines the nature of cosmic evolution during different epochs. A universe with $q<0$ undergoes accelerated expansion, whereas $q>0$ corresponds to a decelerating phase. Further, the condition $q<-1$ signifies a super-accelerated or phantom phase of expansion. From the standard cosmological viewpoint, the values $q=-1$, $q=\tfrac{1}{2}$ and $q=1$ represent the de Sitter, matter-dominated and radiation-dominated epochs, respectively. The redshift-dependent expression for the deceleration parameter is generally written as:
\begin{equation}{\label{32}}
	q(z)= -\frac{\dot{H}}{H^{2}}-1 .
\end{equation} 
The time derivative of the Hubble parameter may be rewritten in terms of the redshift variable as $\dot{H} = -(1+z)H(z)\frac{dH(z)}{dz}$. By substituting Eqs.~(\ref{23}) to (\ref{25}) into Eq.~(\ref{32}), the corresponding expressions for the deceleration parameter $q(z)$ within the considered cosmological models are obtained as follows: 
\begin{equation}{\label{33}}
	q(z)= -1 + \zeta \left[\frac{\left(\eta(1+z)\right)^{3\zeta}-3}{1+\left(\eta(1+z)\right)^{3\zeta}}\right]      \qquad \qquad \qquad \qquad     \text{(for Model-I)}
\end{equation}
\begin{equation}{\label{34}}
	q(z)= -1 + \zeta \left[\frac{\left(\eta(1+z)\right)^{4\zeta}-4}{1+\left(\eta(1+z)\right)^{4\zeta}}\right]       \qquad \qquad \qquad \qquad     \text{(for Model-II)}
\end{equation}
\begin{equation}{\label{35}}
	q(z)= -1 + \zeta \left[\frac{\left(\eta(1+z)\right)^{5\zeta}-5}{1+\left(\eta(1+z)\right)^{5\zeta}}\right]        \qquad \qquad \qquad \qquad     \text{(for Model-III)}
\end{equation}
\begin{figure}[ht]
	\centering
	\includegraphics[width=12cm, height=5cm]{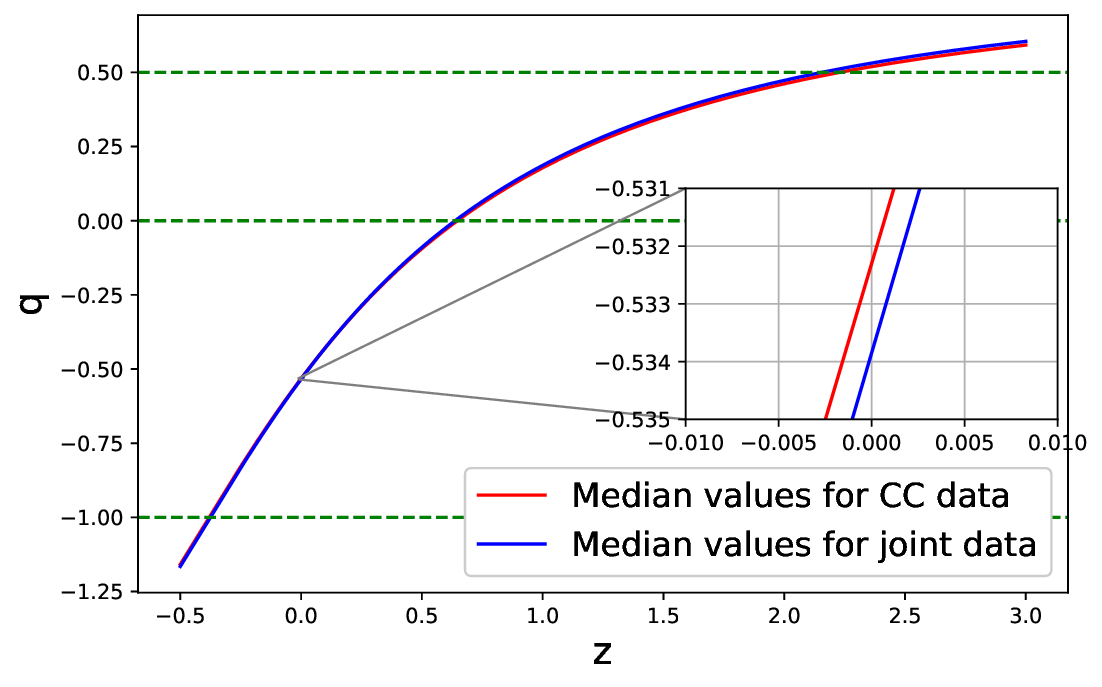}
	\caption{\textbf{For Model-I:} Plot of $q(z)$ with $z$.}
	\label{fig:5}
\end{figure}
\begin{figure}[ht]
	\centering
	\includegraphics[width=12cm, height=5cm]{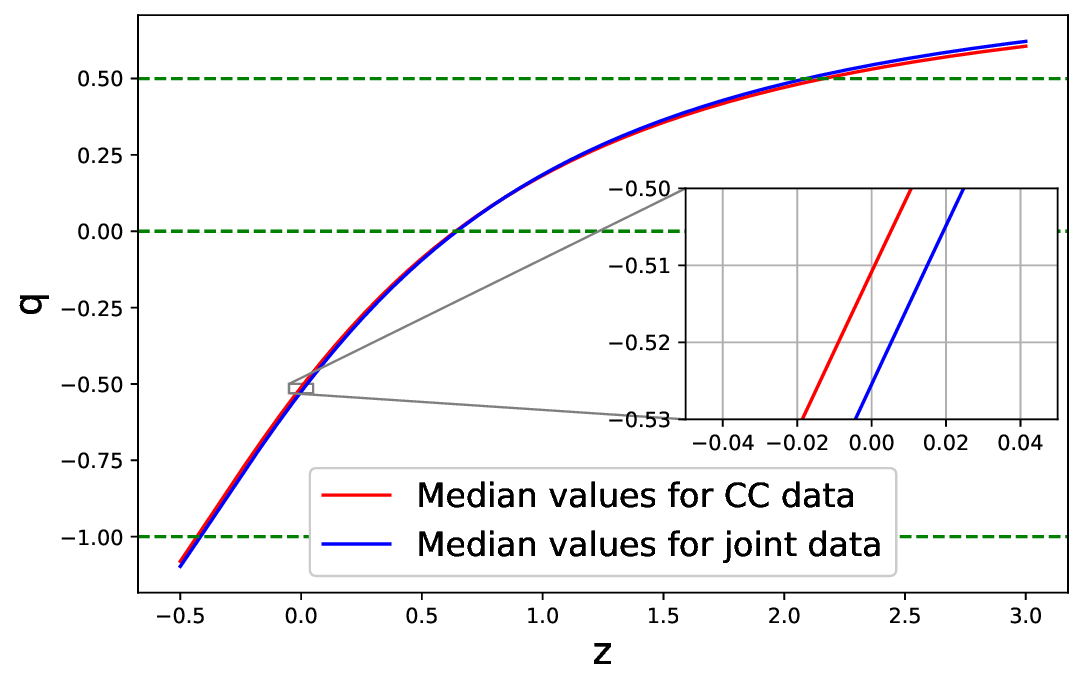}
	\caption{\textbf{For Model-II:} Plot of $q(z)$ with $z$.}
	\label{fig:6}
\end{figure}
\begin{figure}[ht]
	\centering
	\includegraphics[width=12cm, height=5cm]{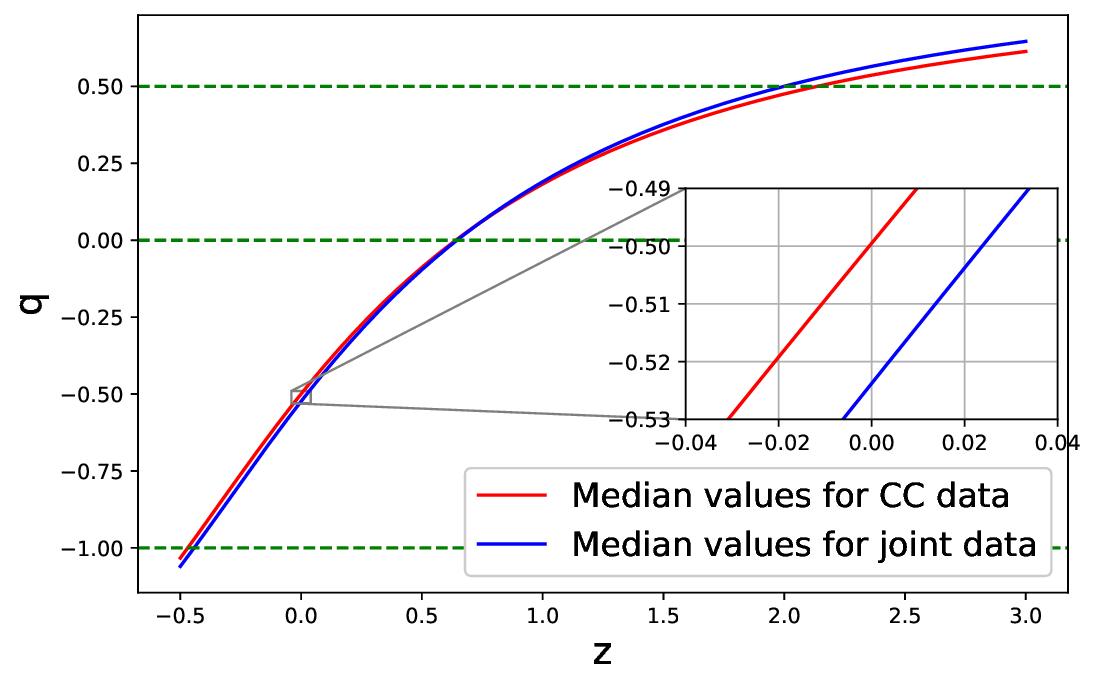}
	\caption{\textbf{For Model-III:} Plot of $q(z)$ with $z$.}
	\label{fig:7}
\end{figure}
\vspace{0.1cm}\\                         
The reconstructed evolution of the deceleration parameter for the CC and combined datasets is displayed in Figures~(\ref{fig:5}) to (\ref{fig:7}). The graphical behavior reveals that the universe evolves from an earlier decelerated expansion stage toward the currently accelerating epoch. By employing the median best-fit values of the model parameters, the present values of the deceleration parameter for Model-I are obtained as $q_{0}=-0.5323$ for the CC dataset and $q_{0}=-0.5338$ for the joint dataset. In the case of Model-II, the corresponding values are $q_{0}=-0.5108$ and $q_{0}=-0.5254$, while Model-III yields $q_{0}=-0.4995$ and $q_{0}=-0.5238$, respectively. The negative sign of the present deceleration parameter in all models indicates that the universe is currently in an accelerated expansion phase. Hence, the obtained results remain fully consistent with the observed late-time cosmic acceleration. It is worth noting that the obtained values of $q_{0}$ are in close agreement with the observationally inferred $\Lambda$CDM prediction $q_{0} \approx -0.55$~\cite{2020A&A...641A...6P}. This consistency indicates that the present models successfully reproduce the observed late-time accelerated expansion behavior, in agreement with standard cosmological observations. At the same time, the models also allow for possible deviations from the standard $\Lambda$CDM scenario at future cosmic epochs, thereby accommodating more general dynamical evolutions such as phantom behavior. The redshift corresponding to the transition from decelerated to accelerated expansion is determined to be $z=0.6547$ and $z=0.6523$ for Model-I using the CC and joint datasets, respectively. Similarly, for Model-II, the transition occurs at $z=0.6442$ (CC) and $z=0.6495$ (joint), while for Model-III it is $z=0.6486$ (CC) and $z=0.6504$ (joint), respectively.
\vspace{0.1cm}\\
To ensure consistency across different cosmological epochs, we now examine the high-redshift and late-time asymptotic behavior of the models. In the high-redshift regime, all three models approach $q = \tfrac{1}{2}$, thereby recovering a matter-dominated past consistent with standard cosmology. At late times, each model undergoes a transition from decelerated to accelerated expansion and may further evolve into a super-accelerated phase driven by phantom-like dark energy behavior. Since this expansion can exceed the de Sitter limit, the effective dark energy sector may be interpreted within the framework of quintom cosmology~\cite{Zhao2006}.
\begin{figure}[!htb]
	\captionsetup{skip=0.4\baselineskip,size=footnotesize}
	\begin{minipage}{0.50\textwidth}
		\centering
		\includegraphics[width=7.6cm,height=6cm]{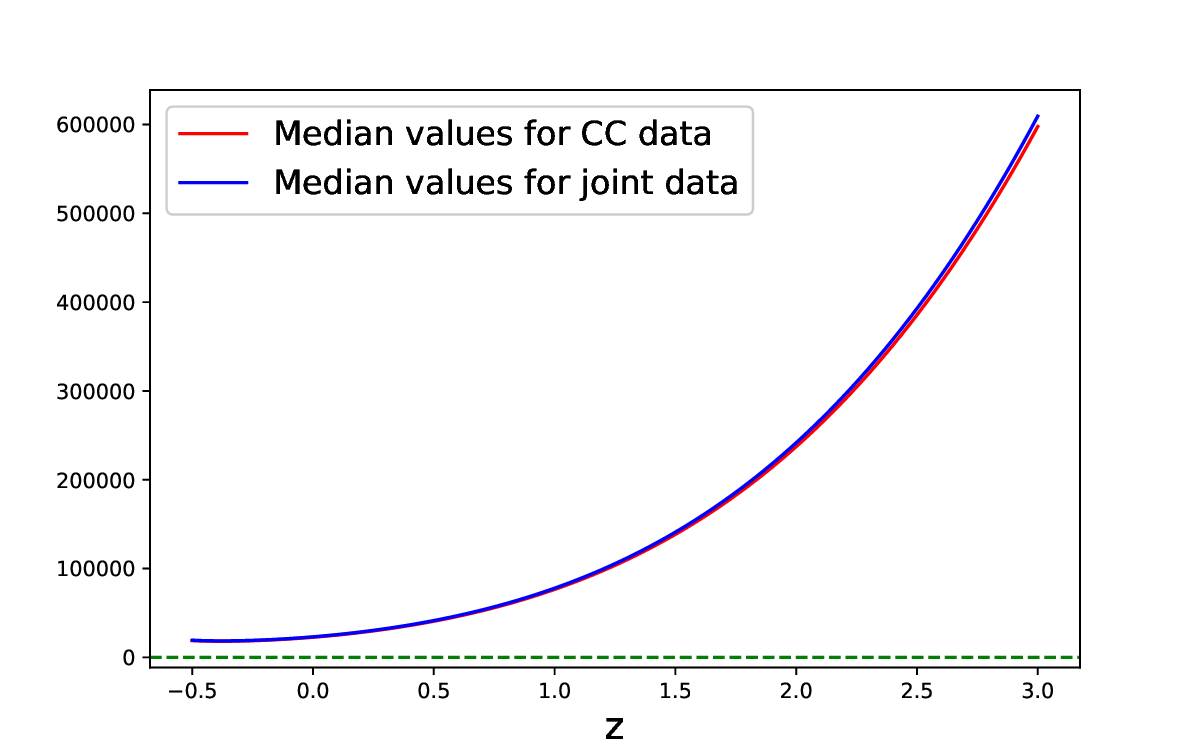}
		\caption{\textbf{For Model-I:} Plot of energy density ($\rho$) with $\mathit{z}$.}
		\label{fig:8}
	\end{minipage}\hfill
	\begin{minipage}{0.50\textwidth}
		\centering
		\includegraphics[width=7.6cm,height=6cm]{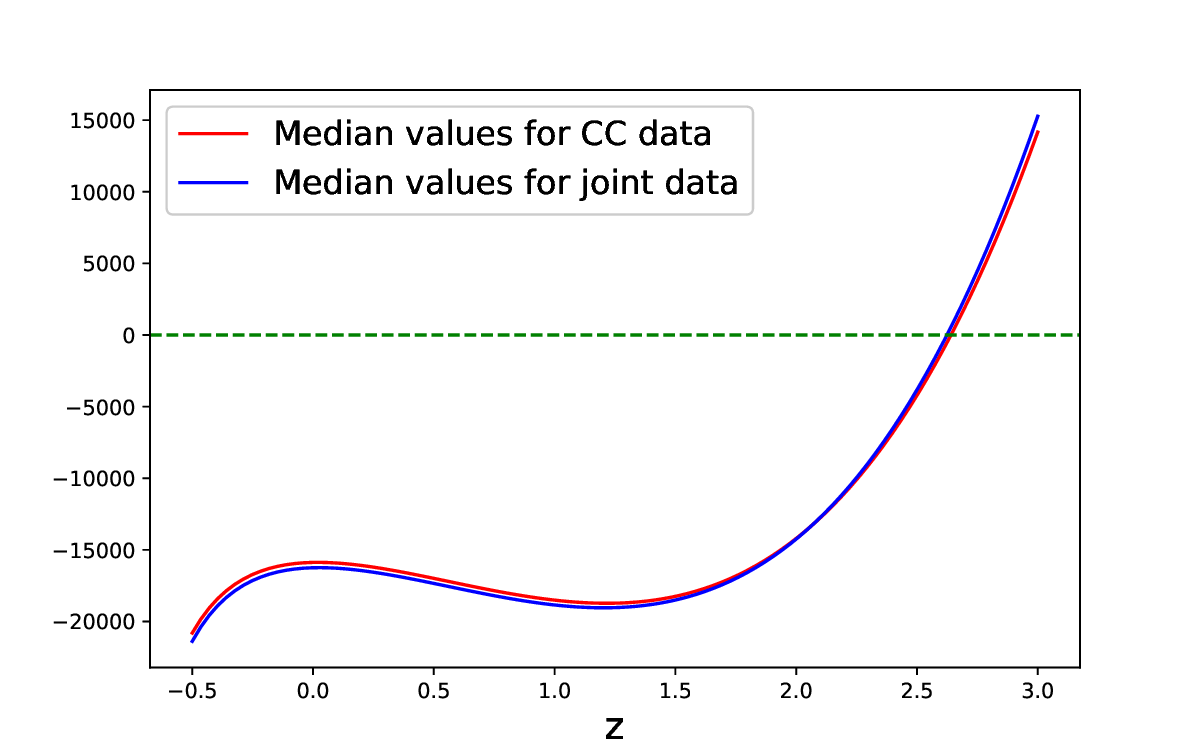}
		\caption{\textbf{For Model-I:} Plot of pressure ($p$) with $\mathit{z}$.}
		\label{fig:9}
	\end{minipage}
\end{figure}
\begin{figure}[!htb]
	\captionsetup{skip=0.4\baselineskip,size=footnotesize}
	\begin{minipage}{0.50\textwidth}
		\centering
		\includegraphics[width=7.6cm,height=6cm]{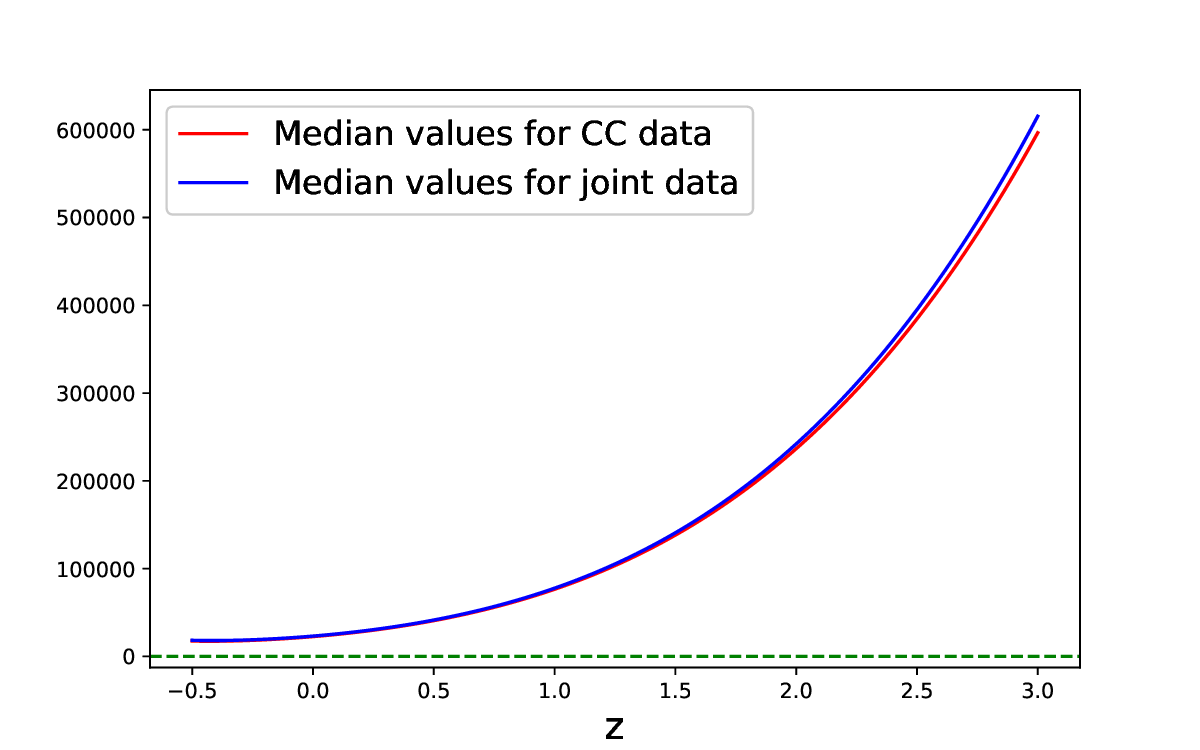}
		\caption{\textbf{For Model-II:} Plot of energy density ($\rho$) with $\mathit{z}$.}
		\label{fig:10}
	\end{minipage}\hfill
	\begin{minipage}{0.50\textwidth}
		\centering
		\includegraphics[width=7.6cm,height=6cm]{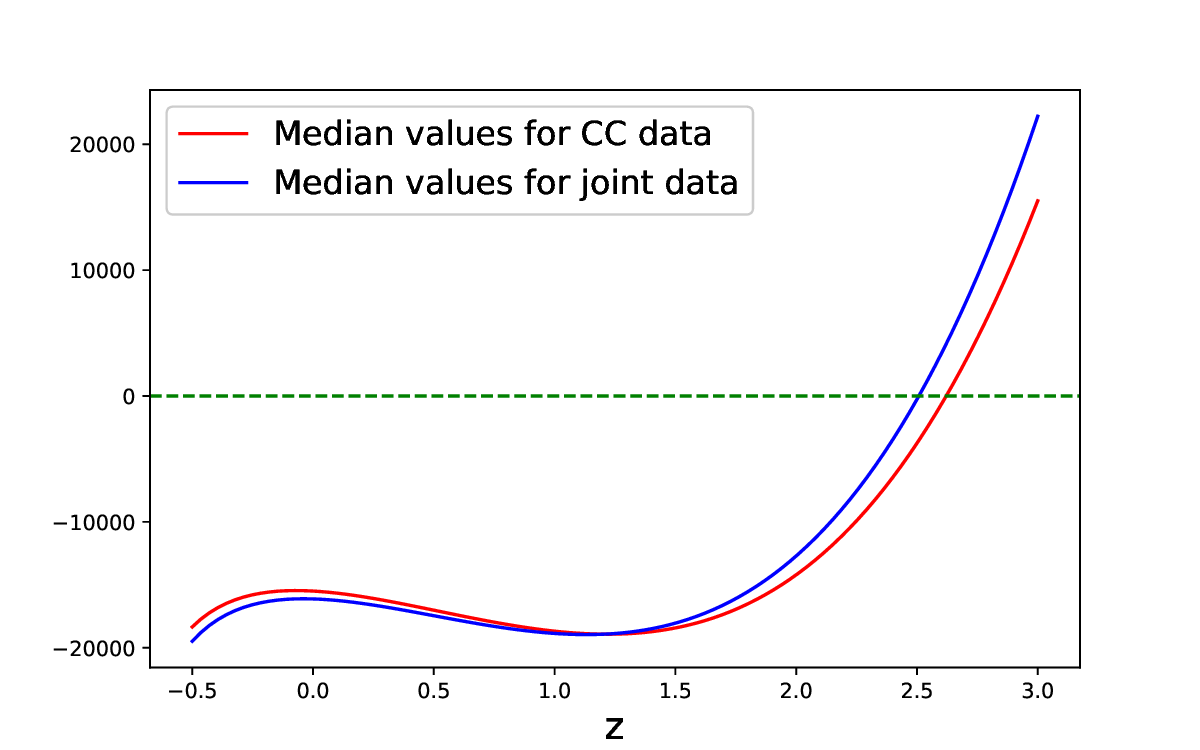}
		\caption{\textbf{For Model-II:} Plot of pressure ($p$) with $\mathit{z}$.}
		\label{fig:11}
	\end{minipage}
\end{figure}
\begin{figure}[!htb]
	\captionsetup{skip=0.4\baselineskip,size=footnotesize}
	\begin{minipage}{0.50\textwidth}
		\centering
		\includegraphics[width=7.6cm,height=6cm]{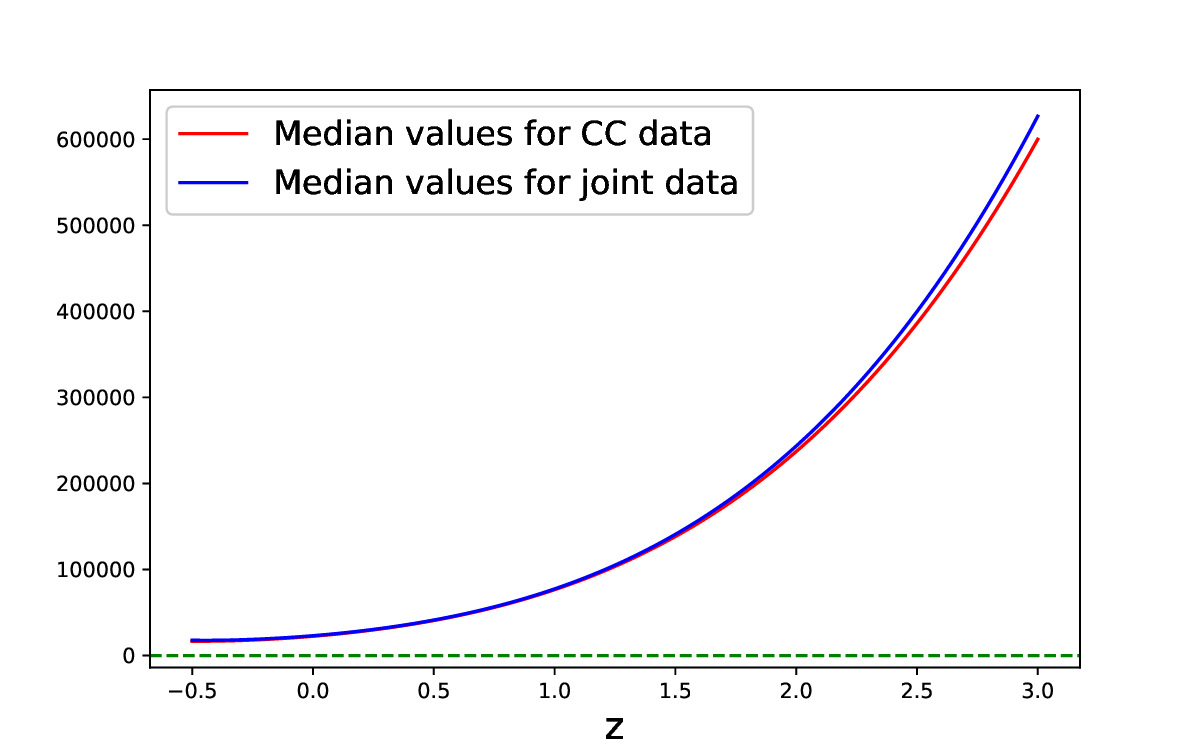}
		\caption{\textbf{For Model-III:} Plot of energy density ($\rho$) with $\mathit{z}$.}
		\label{fig:12}
	\end{minipage}\hfill
	\begin{minipage}{0.50\textwidth}
		\centering
		\includegraphics[width=7.6cm,height=6cm]{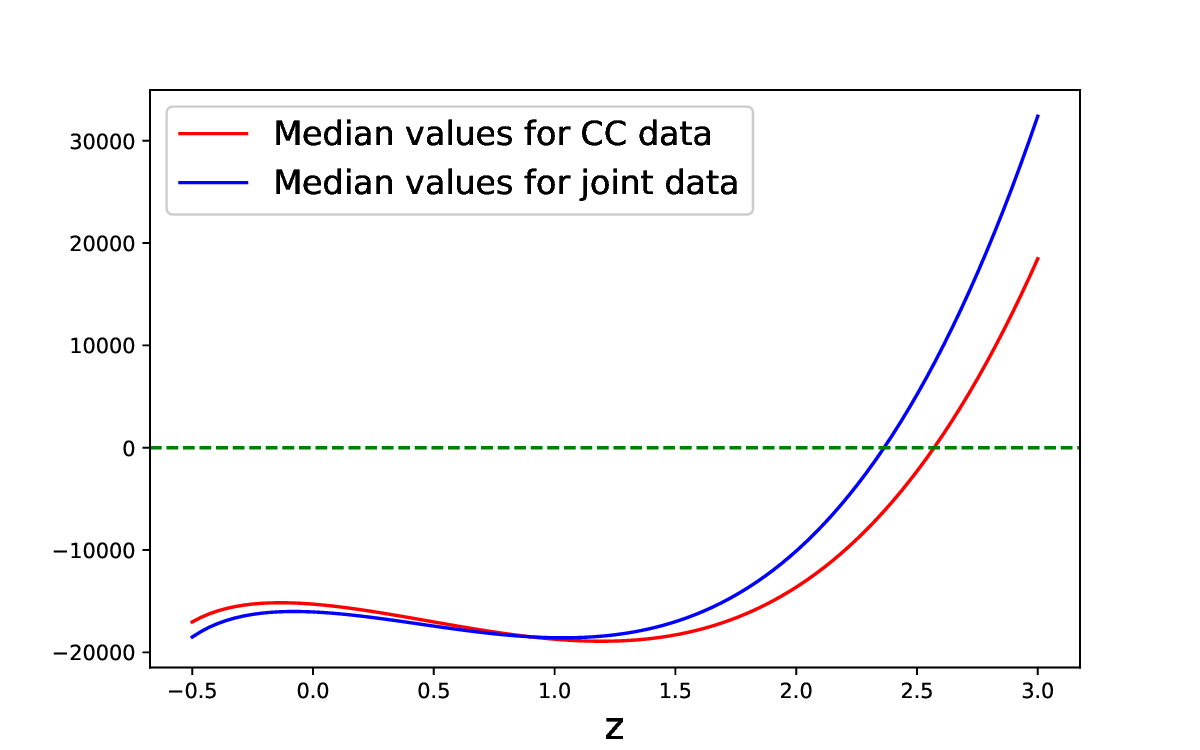}
		\caption{\textbf{For Model-III:} Plot of pressure ($p$) with $\mathit{z}$.}
		\label{fig:13}
	\end{minipage}
\end{figure}
\subsection{Analysis of the energy density, pressure and EoS parameter}\label{subsec:6.2}
\begin{figure}[ht]
	\centering
	\includegraphics[width=12cm, height=5cm]{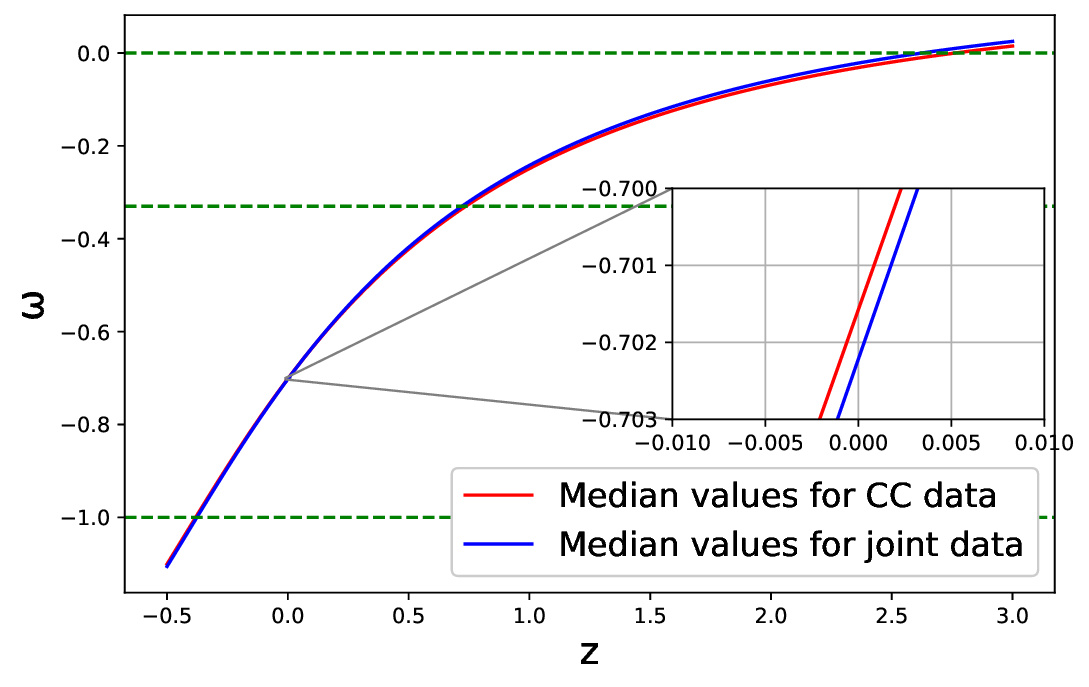}
	\caption{\textbf{For Model-I:} Plot of EoS parameter ($\omega$) with $z$.}
	\label{fig:14}
\end{figure}

\begin{figure}[ht]
	\centering
	\includegraphics[width=12cm, height=5cm]{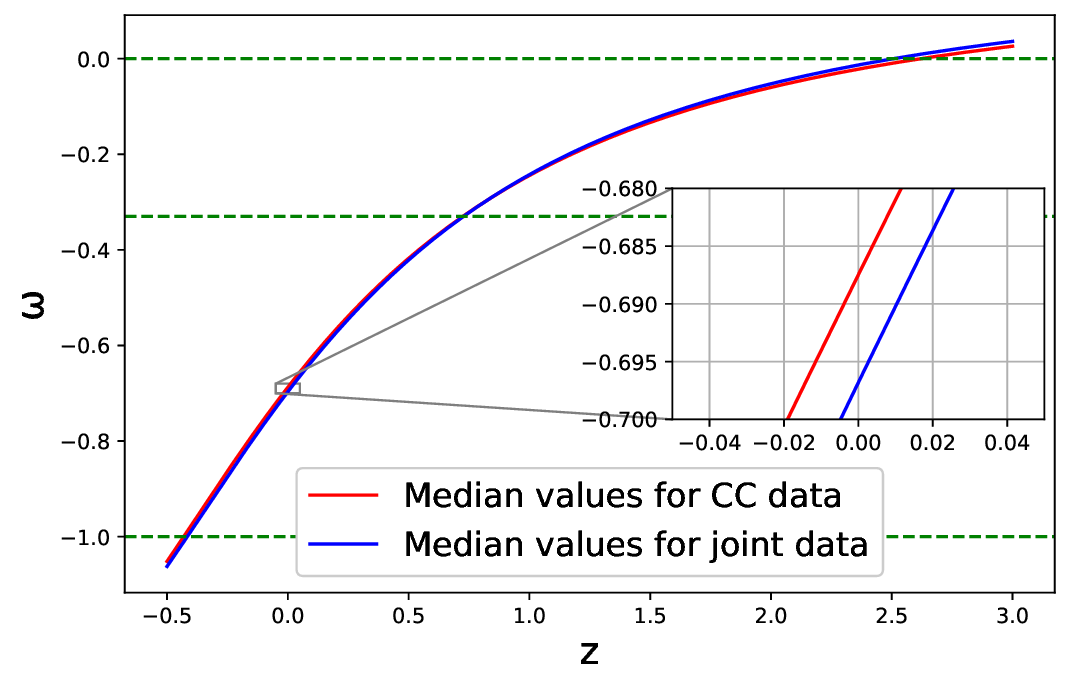}
	\caption{\textbf{For Model-II:} Plot of EoS parameter ($\omega$) with $z$.}
	\label{fig:15}
\end{figure}

\begin{figure}[ht]
	\centering
	\includegraphics[width=12cm, height=5cm]{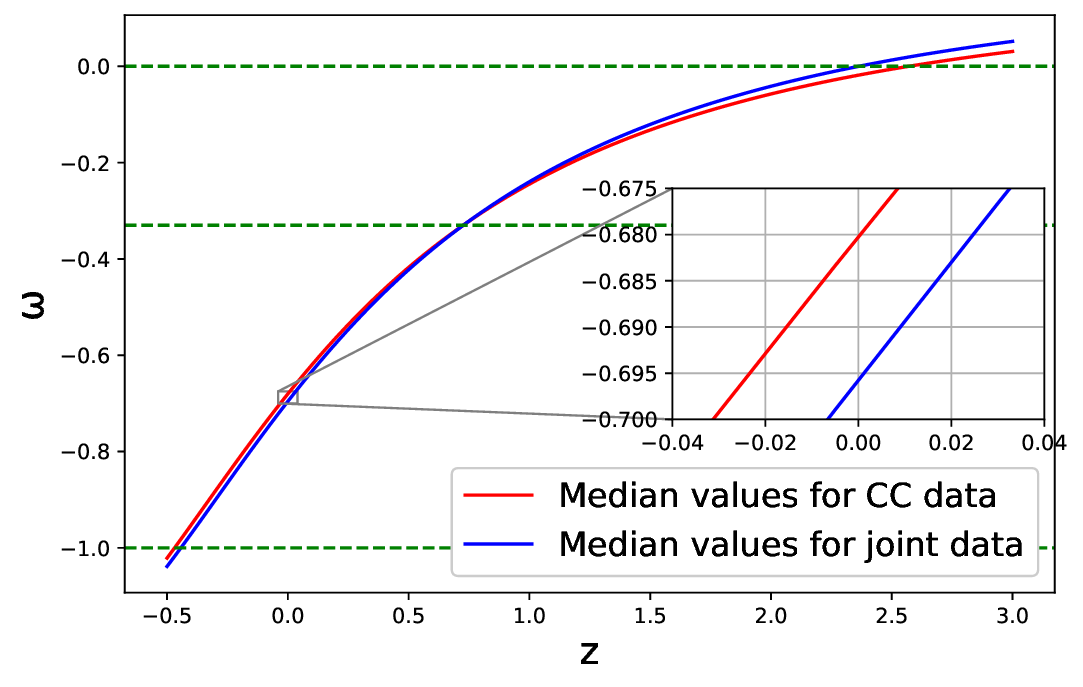}
	\caption{\textbf{For Model-III:} Plot of EoS parameter ($\omega$) with $z$.}
	\label{fig:16}
\end{figure}

The present study examines the evolution of fundamental cosmological quantities, particularly the energy density and pressure. Using the constrained parameter space, it is observed that the energy density preserves its positive nature throughout the cosmic history, while the pressure exhibits a transition from positive values to negative values at recent epochs. As cosmic evolution proceeds from a decelerated phase toward accelerated expansion regime, the positivity of the energy density is preserved, whereas the negative pressure emerges due to the increasing influence of dark energy. This behavior aligns well with the standard expectations for a late-time accelerating universe.
\vspace{0.1cm}\\
Using Eqs.~(\ref{16}), (\ref{17}) and (\ref{23}), the corresponding expressions for the energy density and pressure associated with Model-I can be derived as follows:
\begin{equation}{\label{36}}
\rho(z) =\left(\frac{3}{2\lambda-1}\right)^{1/\lambda} \times \left\{H_{0}^{2}(1+\eta^{3\zeta})^{\frac{-8}{3}} (1+z)^{-6\zeta} \left[1+(\eta (1+z))^{3\zeta} \right]^{\frac{8}{3}}\right\}^{1/\lambda}     \qquad\qquad    \text{(for Model-I)}
\end{equation}
\begin{equation} {\label{37}}
\resizebox{1.0\textwidth}{!}{$		
p(z)= \frac{-1}{3\lambda}\left(\frac{3}{2\lambda-1}\right)^{1/\lambda} \times \left(3\lambda+\frac{\zeta(4\lambda-2)\left(3-\left(\eta(1+z)\right)^{3\zeta}\right)}{1+\left(\eta(1+z)\right)^{3\zeta}}\right) \times \left\{H_{0}^{2}(1+\eta^{3\zeta})^{\frac{-8}{3}} (1+z)^{-6\zeta} \left[1+(\eta (1+z))^{3\zeta} \right]^{\frac{8}{3}}\right\}^{1/\lambda}   \quad \text{(for Model-I)}
$}              
\end{equation} 
By combining Eqs.~(\ref{16}), (\ref{17}) and (\ref{24}), the corresponding expressions for the energy density and pressure in Model-II can be written as follows:
\begin{equation}{\label{38}}
	\rho(z) =\left(\frac{3}{2\lambda-1}\right)^{1/\lambda} \times \left\{H_{0}^{2}(1+\eta^{4\zeta})^{\frac{-5}{2}} (1+z)^{-8\zeta} \left[1+(\eta (1+z))^{4\zeta} \right]^{\frac{5}{2}}\right\}^{1/\lambda}     \qquad\qquad    \text{(for Model-II)}
\end{equation}
\begin{equation} {\label{39}}
	\resizebox{1.0\textwidth}{!}{$		
		p(z)= \frac{-1}{3\lambda}\left(\frac{3}{2\lambda-1}\right)^{1/\lambda} \times \left(3\lambda+\frac{\zeta(4\lambda-2)\left(4-\left(\eta(1+z)\right)^{4\zeta}\right)}{1+\left(\eta(1+z)\right)^{4\zeta}}\right) \times \left\{H_{0}^{2}(1+\eta^{4\zeta})^{\frac{-5}{2}} (1+z)^{-8\zeta} \left[1+(\eta (1+z))^{4\zeta} \right]^{\frac{5}{2}}\right\}^{1/\lambda}   \quad \text{(for Model-II)}
		$}              
\end{equation} 
Further, using Eqs.~(\ref{16}), (\ref{17}) and (\ref{25}), the corresponding expressions for the energy density and pressure associated with Model-III can be derived as follows:
\begin{equation}{\label{40}}
	\rho(z) =\left(\frac{3}{2\lambda-1}\right)^{1/\lambda} \times \left\{H_{0}^{2}(1+\eta^{5\zeta})^{\frac{-12}{5}} (1+z)^{-10\zeta} \left[1+(\eta (1+z))^{5\zeta} \right]^{\frac{12}{5}}\right\}^{1/\lambda}     \qquad\qquad    \text{(for Model-III)}
\end{equation}
\begin{equation} {\label{41}}
	\resizebox{1.0\textwidth}{!}{$		
		p(z)= \frac{-1}{3\lambda}\left(\frac{3}{2\lambda-1}\right)^{1/\lambda} \times \left(3\lambda+\frac{\zeta(4\lambda-2)\left(5-\left(\eta(1+z)\right)^{5\zeta}\right)}{1+\left(\eta(1+z)\right)^{5\zeta}}\right) \times \left\{H_{0}^{2}(1+\eta^{5\zeta})^{\frac{-12}{5}} (1+z)^{-10\zeta} \left[1+(\eta (1+z))^{5\zeta} \right]^{\frac{12}{5}}\right\}^{1/\lambda}   \quad \text{(for Model-III)}
		$}              
\end{equation} 
The evolutionary behavior of the energy density and pressure is illustrated in Figures~(\ref{fig:8}) to (\ref{fig:13}) for Models I to III. For the observationally constrained values of the model parameters, the energy density exhibits a well-defined monotonic increase with redshift ($z$) (corresponding to a decreasing trend with cosmic time ($t$)), while remaining strictly positive throughout the entire cosmic evolution. This persistent positivity of the energy density not only ensures the physical admissibility of the models but also reflects the sustained contribution of the cosmic fluid in driving the expansion dynamics of the universe.
In contrast, the pressure undergoes a significant transition and attains negative values in the recent cosmic past, continuing to remain negative at the present epoch. Such negative pressure is a key characteristic of dark energy and plays a crucial role in generating the observed accelerated expansion of the universe by effectively counteracting gravitational attraction. The persistence of negative pressure at late times indicates the dominance of a repulsive component in the cosmic energy budget, thereby leading to an accelerated expansion phase.
\vspace{0.1cm}\\
From a physical perspective, the combined behavior of positive energy density and negative pressure provides strong support for a dark energy dominated universe at late times. Moreover, the well-behaved evolution of these quantities across different redshift regimes demonstrates the robustness of the proposed models in capturing both the early decelerated phase and the subsequent accelerated expansion. These results are fully consistent with current cosmological observations and reinforce the viability of the models as plausible candidates for describing the late-time dynamics of the universe. For the purpose of graphical illustration, we adopt the parameter value $\lambda = 0.96$.
The equation of state (EoS) parameter is one of the key quantities used to describe the dynamical properties of dark energy in cosmology. It connects the cosmic pressure $p$ with the corresponding energy density $\rho$ through the relation $\omega=\frac{p}{\rho}$. The numerical value of $\omega$ determines the dominant evolutionary behavior of the universe during different cosmic epochs. In particular, the value $\omega=0$ represents a pressureless matter dominated universe, whereas $\omega=\tfrac{1}{3}$ corresponds to the radiation era. The case $\omega=-1$ is associated with vacuum energy and the de Sitter phase of expansion. Furthermore, the requirement for an accelerating universe is satisfied when $\omega<-\tfrac{1}{3}$. This interval includes both the quintessence regime ($-1<\omega<-\tfrac{1}{3}$) and the phantom regime ($\omega<-1$), which are commonly used to explain different dark energy behaviors responsible for the observed late-time acceleration of the universe.
\vspace{0.2cm}\\
Using Eqs.~(\ref{18}) and (\ref{23}), the EoS parameter for Model-I is obtained as:
\begin{equation} {\label{42}}	
\omega(z)= -1+\left(\frac{1}{3\lambda}\right) \times \left(\frac{\zeta(2-4\lambda)\left(3-\left(\eta(1+z)\right)^{3\zeta}\right)}{1+\left(\eta(1+z)\right)^{3\zeta}}\right) \quad \qquad \text{(for Model-I)}             
\end{equation} 
From Eqs.~(\ref{18}) and (\ref{24}), the EoS parameter for Model-II is derived as:
\begin{equation} {\label{43}}	
	\omega(z)= -1+\left(\frac{1}{3\lambda}\right) \times \left(\frac{\zeta(2-4\lambda)\left(4-\left(\eta(1+z)\right)^{4\zeta}\right)}{1+\left(\eta(1+z)\right)^{4\zeta}}\right) \quad \qquad \text{(for Model-II)}             
\end{equation} 
Further, using Eqs.~(\ref{18}) and (\ref{25}), the EoS parameter for Model-III is obtained as:
\begin{equation} {\label{44}}	
	\omega(z)= -1+\left(\frac{1}{3\lambda}\right) \times \left(\frac{\zeta(2-4\lambda)\left(5-\left(\eta(1+z)\right)^{5\zeta}\right)}{1+\left(\eta(1+z)\right)^{5\zeta}}\right) \quad \qquad \text{(for Model-III)}             
\end{equation} 
Figures~(\ref{fig:14}) to (\ref{fig:16}) display the redshift dependence of the EoS parameter $\omega(z)$ for the proposed cosmological models. Using the median parameter estimates, the present-day value $(z=0)$ of the EoS parameter for Model-I is found to be $\omega=-0.7015$ for the CC dataset and $\omega=-0.7022$ for the joint dataset. Similarly, Model-II predicts $\omega=-0.6875$ and $\omega=-0.6968$, while Model-III yields $\omega=-0.6802$ and $\omega=-0.6957$ for the CC and combined observational datasets, respectively. These values consistently lie within the quintessence regime ($-1 < \omega < -\tfrac{1}{3}$), indicating that the present cosmic acceleration is driven by a dynamical dark energy component rather than a pure cosmological constant. A detailed inspection of the evolutionary behavior of $\omega(z)$ reveals that all three models exhibit a coherent and well-defined evolution across the redshift range considered. In particular, the EoS parameter evolves from values closer to the quintessence regime at intermediate redshifts toward more negative values at late times. This trend reflects the increasing dominance of dark energy in the cosmic energy budget, which progressively overcomes the gravitational influence of matter and drives the accelerated expansion of the universe.
\vspace{0.1cm}\\
Furthermore, the trajectories of $\omega(z)$ suggest that, for the median values of the model parameters, the EoS parameter can cross the cosmological constant boundary ($\omega = -1$) and enter the phantom regime in the late-time limit. Such a transition is of significant cosmological relevance, given that it points toward a dynamical dark energy model capable of evolving beyond the standard $\Lambda$CDM framework. The emergence of phantom-like behavior implies a stronger negative pressure component, which can lead to an enhanced rate of the expanding universe and may have profound implications for the future evolution of the cosmos, including the possibility of a super-accelerated phase. Altogether, the variation of $\omega(z)$ across all three models demonstrates their robustness in capturing both the present acceleration and the potential future dynamics of the universe within a unified framework.
\subsection{Analysis of energy conditions}\label{sec:6.3}
\begin{figure}[ht]
	\centering
	\includegraphics[width=12cm, height=5cm]{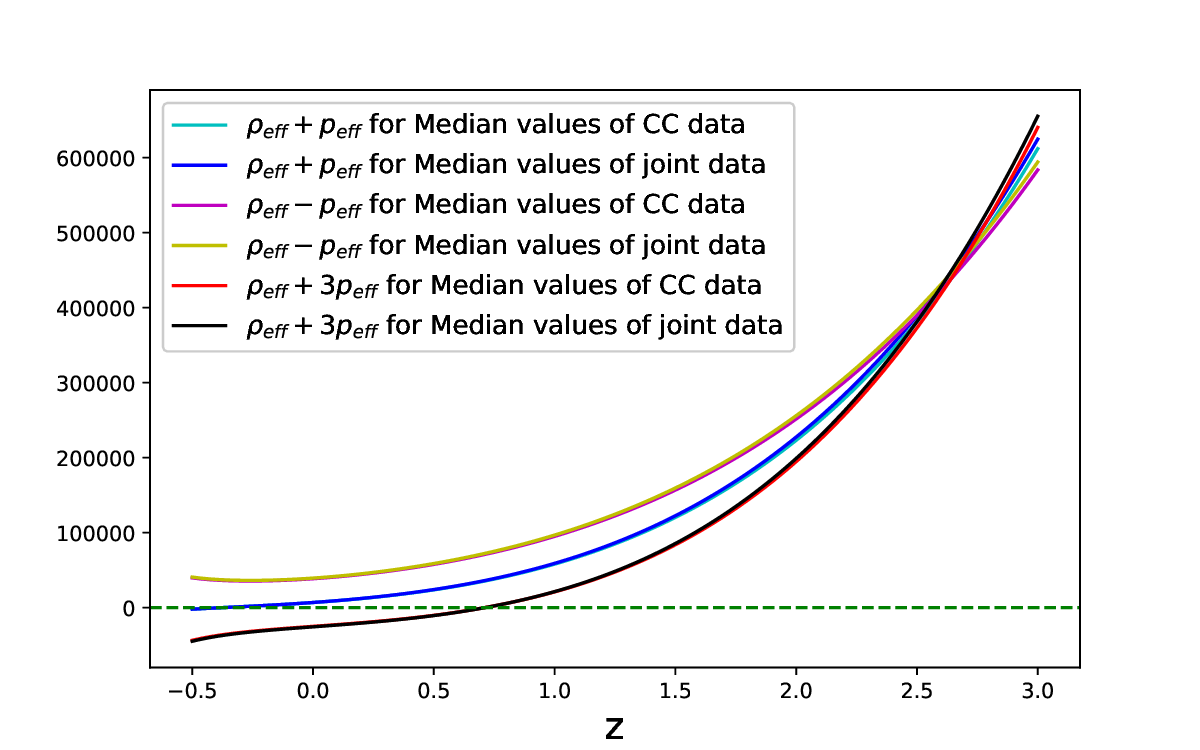}
	\caption{\textbf{For Model-I:} Plot of the evolution of energy condition components with $z$.}
	\label{fig:17}
\end{figure}

\begin{figure}[ht]
	\centering
	\includegraphics[width=12cm, height=5cm]{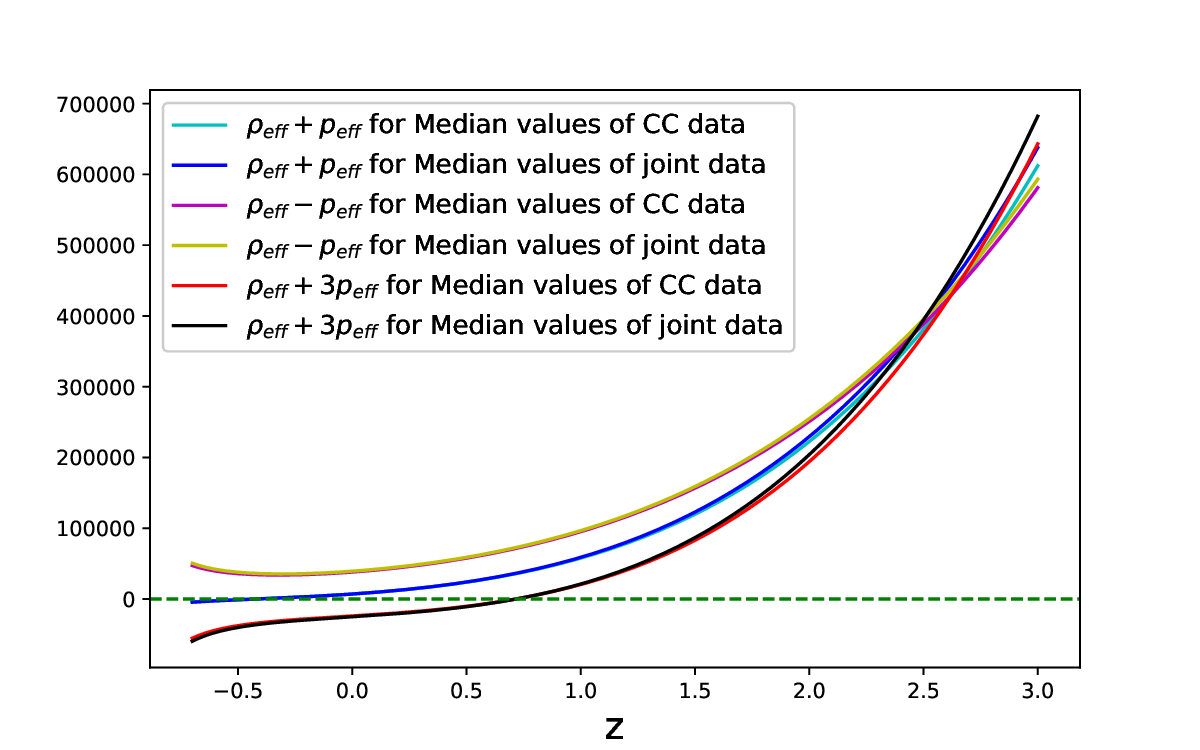}
	\caption{\textbf{For Model-II:} Plot of the evolution of energy condition components with $z$.}
	\label{fig:18}
\end{figure}

\begin{figure}[ht]
	\centering
	\includegraphics[width=12cm, height=5cm]{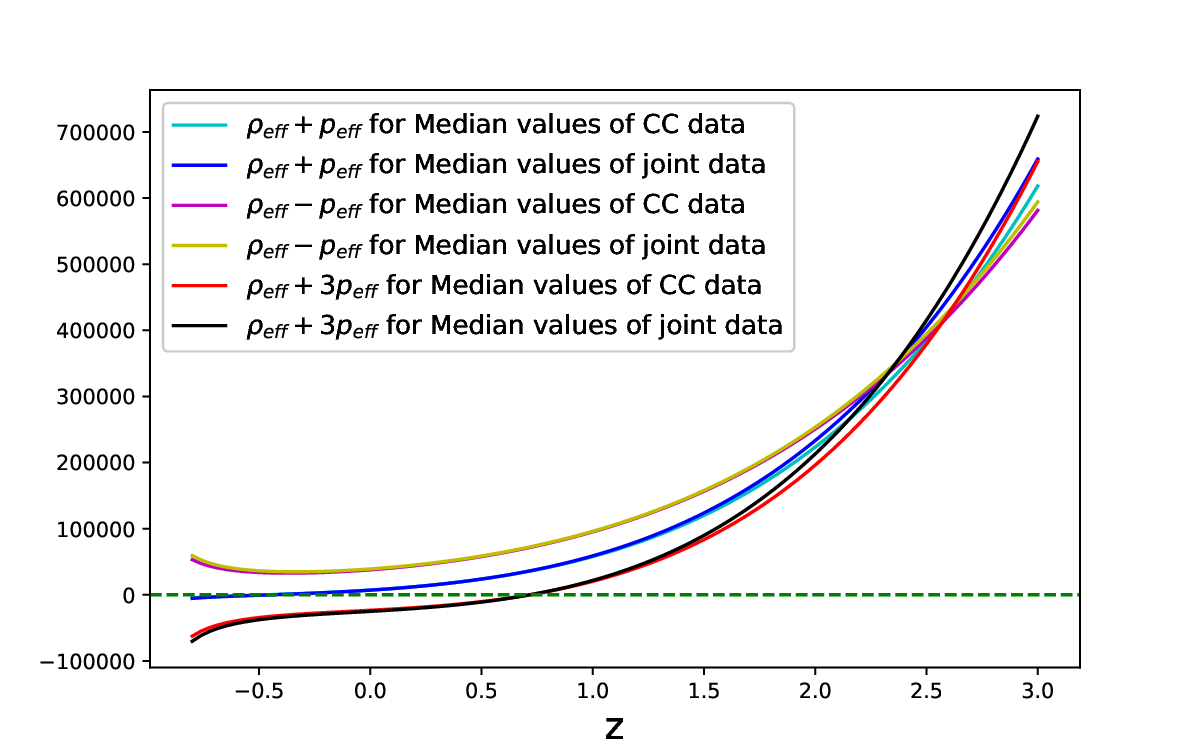}
	\caption{\textbf{For Model-III:} Plot of the evolution of energy condition components with $z$.}
	\label{fig:19}
\end{figure}

The point-wise energy conditions, defined at a given spacetime point and governed solely by the stress energy tensor, can be written as~\cite{visser1997energy,lalke2024cosmic,singh2022lagrangian}:
\begin{itemize}
    \item \textbf{NEC:} The null energy condition is satisfied when the relation $\rho_{\mathrm{eff}} + p_{\mathrm{eff}} \geq 0$ holds, indicating a non-negative combination of the effective energy density and pressure.
    
    \item \textbf{WEC:} The weak energy condition requires both $\rho_{\mathrm{eff}} \geq 0$ and $\rho_{\mathrm{eff}} + p_{\mathrm{eff}} \geq 0$, implying that the effective energy density remains positive.
    
    \item \textbf{DEC:} The dominant energy condition is fulfilled if $\rho_{\mathrm{eff}} \geq |p_{\mathrm{eff}}|$, implying that the effective energy density exceeds the magnitude of the effective pressure.
    
    \item \textbf{SEC:} The strong energy condition holds when the inequalities $\rho_{\mathrm{eff}} + p_{\mathrm{eff}} \geq 0$ and $\rho_{\mathrm{eff}} + 3p_{\mathrm{eff}} \geq 0$ are simultaneously satisfied.
	
\end{itemize}
The SEC, characterized by the inequality $\rho_{\mathrm{eff}} + 3p_{\mathrm{eff}} \geq 0$, is intimately connected to the Raychaudhuri equation and plays a central role in governing the dynamical evolution of the universe~\cite{mishra2025cosmological}. Within the standard cosmological framework, a positive value of the effective gravitational mass term $(\rho_{\mathrm{eff}} + 3p_{\mathrm{eff}})$ corresponds to a decelerating expansion. However, accumulating observational evidence suggests that this condition is violated during the transition from the epoch of structure formation to the present accelerated phase of the universe. The violation of the SEC ($\rho_{\mathrm{eff}} + 3p_{\mathrm{eff}} < 0$), gives rise to an effective repulsive gravitational component, thereby driving the late-time cosmic acceleration. This behavior indicates the presence of a fluid with sufficiently negative pressure, consistent with the phenomenology of dark energy. It is noteworthy that the SEC comprises two independent inequalities and the violation of either is sufficient to signal a breakdown of the condition~\cite{mishra2025cosmological,myrzakulov2023quintessence}.
\vspace{0.1cm}\\
Figures~(\ref{fig:17}) to (\ref{fig:19}) depict the evolution of the energy conditions for Models I to III. The results show that all three models satisfy the NEC, WEC and DEC up to the present epoch, thereby ensuring their physical admissibility within this regime. However, the onset of accelerated expansion necessarily requires a violation of the SEC (particularly the condition $\rho_{\mathrm{eff}} + 3p_{\mathrm{eff}} \geq 0$). As the cosmological dynamics approach the phantom region, the NEC eventually breaks down, which may further lead to violations of the WEC and DEC. The violation of $\rho_{\mathrm{eff}} + p_{\mathrm{eff}} \geq 0$, may serve as a clear signature of phantom-like dark energy behavior. This indicates that a phantom component may play a significant role in the late-time dynamics within the present framework. Moreover, the sustained violation of $\rho_{\mathrm{eff}} + 3p_{\mathrm{eff}} \geq 0$ during the phantom-dominated phase reinforces the persistence of accelerated expansion driven by an effective negative pressure. Taken together, these results are fully consistent with the reconstructed EoS behavior and provide a coherent description of the late-time accelerated expansion of the universe.
\subsection{Statefinder diagnostic analysis}\label{sec:6.4}
\begin{figure}[ht]
	\centering
	\includegraphics[width=10cm, height=5cm]{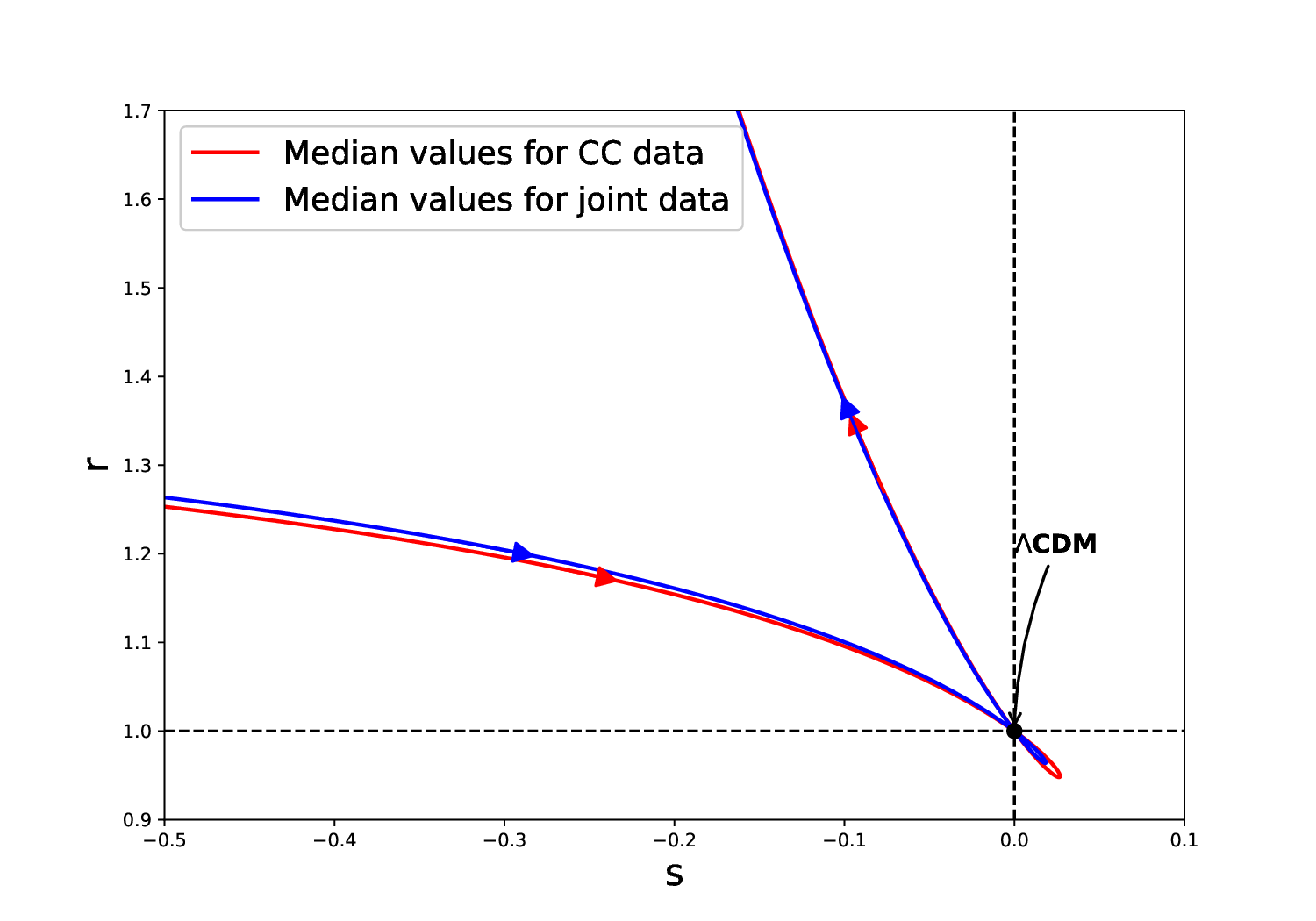}
	\caption{\textbf{For Model-I:} Plot of $s$ and $r$ plane.}
	\label{fig:20}
\end{figure}

\begin{figure}[ht]
	\centering
	\includegraphics[width=10cm, height=5cm]{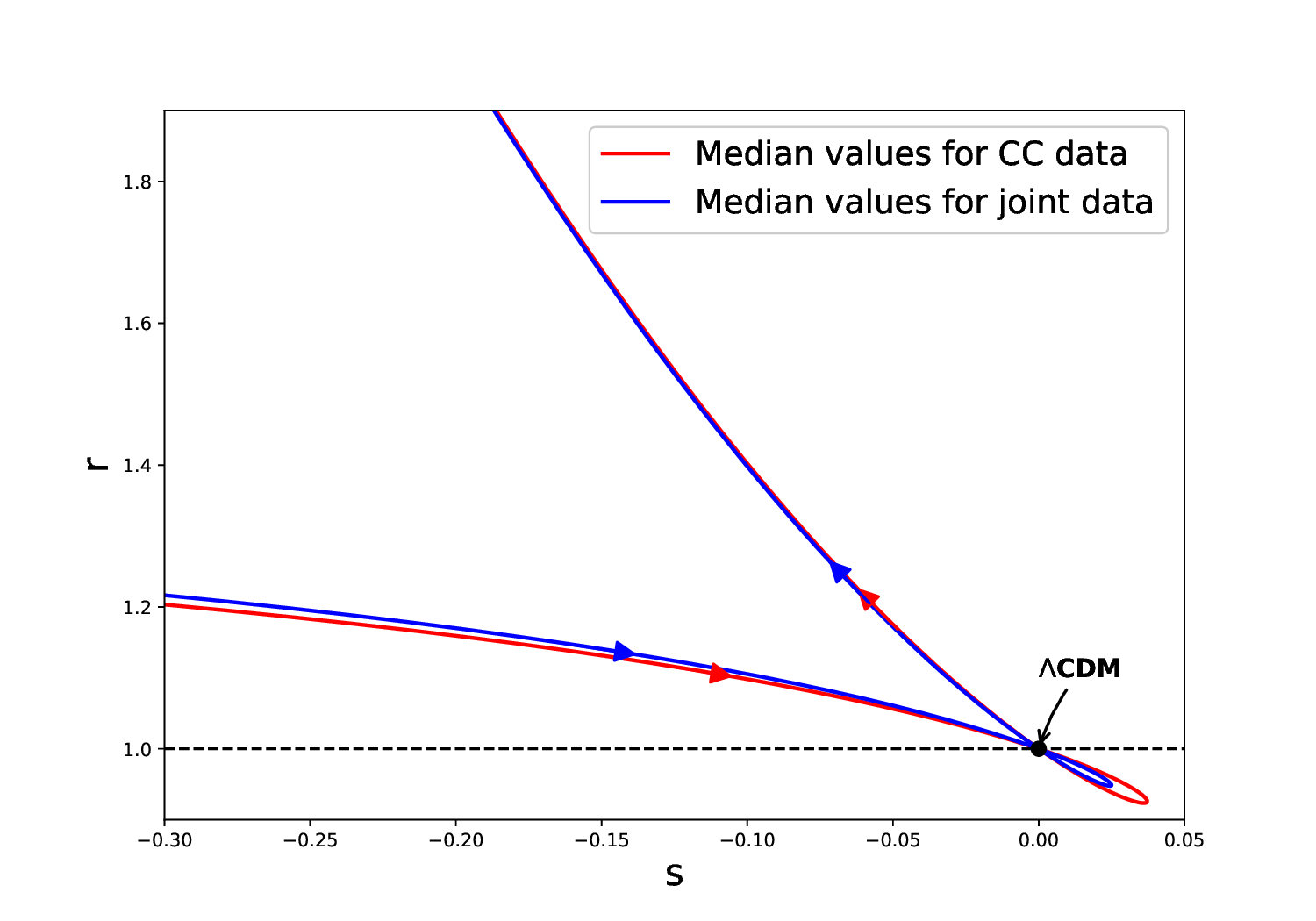}
	\caption{\textbf{For Model-II:} Plot of $s$ and $r$ plane.}
	\label{fig:21}
\end{figure}

\begin{figure}[ht]
	\centering
	\includegraphics[width=10cm, height=5cm]{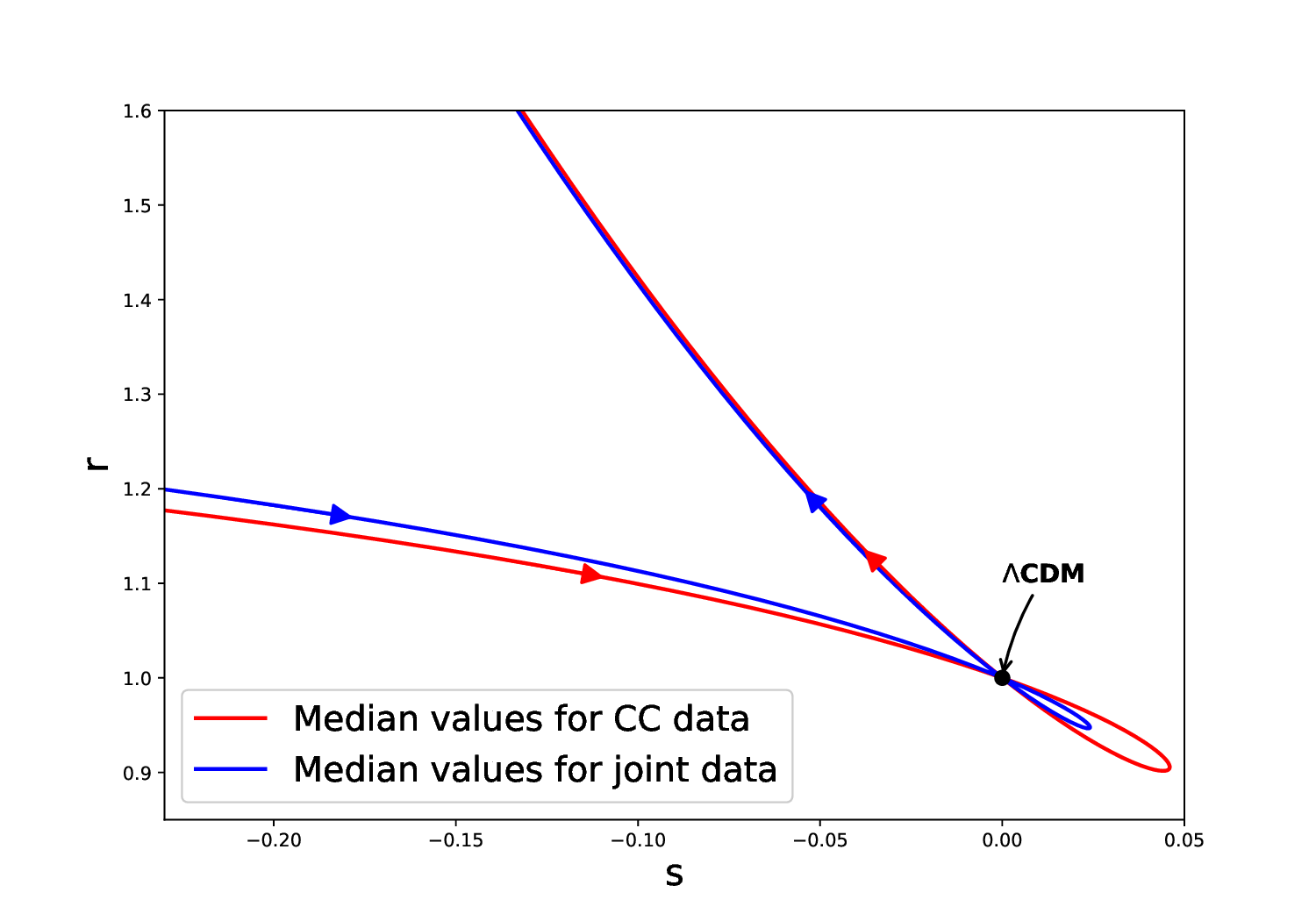}
	\caption{\textbf{For Model-III:} Plot of $s$ and $r$ plane.}
	\label{fig:22}
\end{figure}

Geometrical parameters are widely employed in cosmology to describe the evolutionary behavior of the universe. However, the investigation of dark energy scenarios beyond the conventional $\Lambda$CDM framework requires additional geometrical diagnostics extending beyond the usual parameters $H$ and $q$. For this purpose, higher-order derivatives of the scale factor $a(t)$ become particularly important, since they provide deeper information about the expansion history and help differentiate between various cosmological models. In this direction, the statefinder diagnostic characterized by the parameter pair $\left\{r, s\right\}$~\cite{Sahni2003} has been introduced as an effective geometrical approach for studying the dynamics of dark energy models. It facilitates a systematic classification of dark energy models by examining their evolutionary trajectories in the $\left\{r, s\right\}$ plane. The statefinder parameters $\left\{r, s\right\}$ are defined as follows
\begin{equation}{\label{45}}
	r=\frac{\dddot a}{aH^{3}}= q+2q^{2}+(1+z)\frac{dq}{dz},
\end{equation}
\begin{equation}{\label{46}}
	s= \frac{r-1}{3(q-\frac{1}{2})}, \quad \text{where} \quad q\neq \frac{1}{2}.
\end{equation}
The evolutionary dynamics of different dark energy models discussed in the literature can be efficiently characterized through the statefinder pair $\left\{r, s\right\}$:
\begin{itemize}
	\item In the Chaplygin gas (CG) scenario, the statefinder parameters generally satisfy $(r>1,\; s<0)$.
	
	\item The standard $\Lambda$CDM cosmology is represented by the fixed point $(r=1,\; s=0)$.
	
	\item In the Quintessence model, the parameters usually follow $(r<1,\; s>0)$.
	
	\item For the holographic dark energy (HDE) model, the statefinder pair is given by $(r=1,\; s=\frac{2}{3})$.
	
	\item The standard cold dark matter (SCDM) model corresponds to $(r=1,\; s=1)$.
	
\end{itemize}
The statefinder trajectories in the $\left\{r, s\right\}$ plane for Model-I, Model-II and Model-III are depicted in Figures~(\ref{fig:20}) to (\ref{fig:22}). For all three models, the trajectories originate in the Chaplygin gas regime at early times, pass through the $\Lambda$CDM fixed point and asymptotically approach a unified dark sector behavior at late times. This characteristic evolution demonstrates the robustness of the proposed models in seamlessly interpolating between distinct cosmological phases, thereby capturing a smooth transition from an early Chaplygin gas dominated era to a late-time unified dark matter-dark energy scenario. The observed statefinder dynamics agree well with previously reported studies~\cite{fei2013statefinder}, further reinforcing the physical viability of the models.
\begin{figure}[!htb]
	\captionsetup{skip=0.4\baselineskip,size=footnotesize}
	\begin{minipage}{0.50\textwidth}
		\centering
		\includegraphics[width=7.6cm,height=6cm]{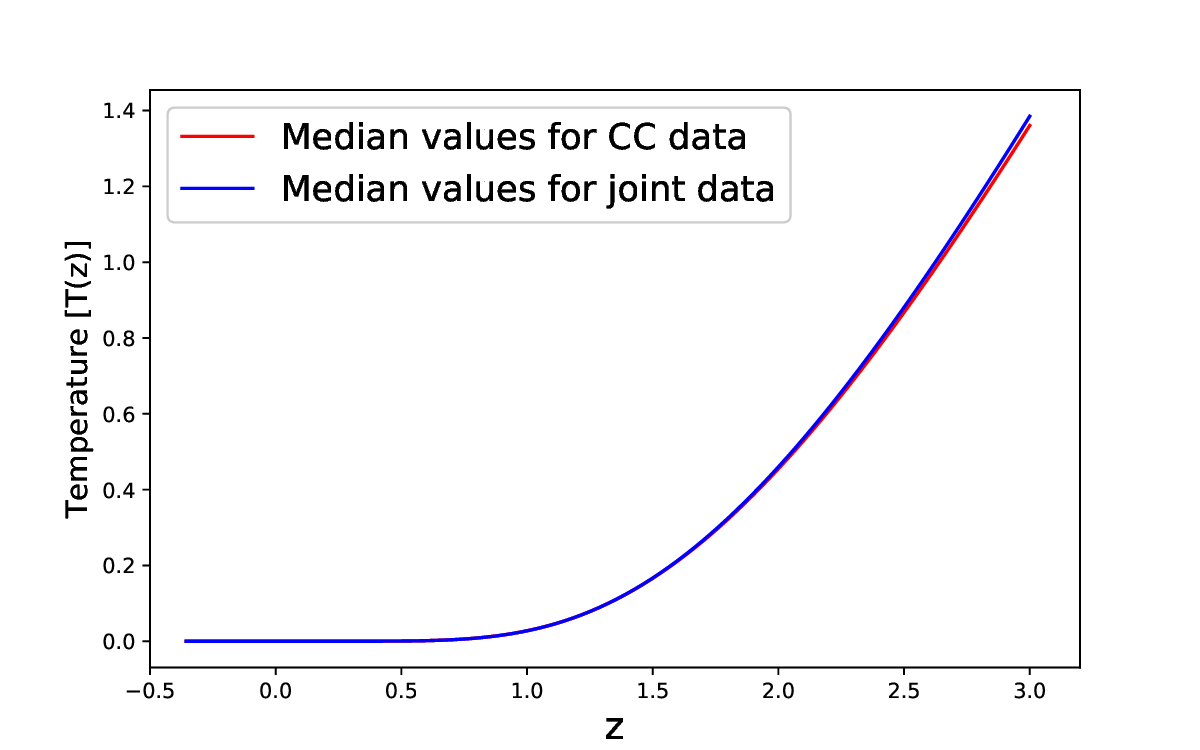}
		\caption{\textbf{For Model-I:} Plot of temperature with $\mathit{z}$.}
		\label{fig:23}
	\end{minipage}\hfill
	\begin{minipage}{0.50\textwidth}
		\centering
		\includegraphics[width=7.6cm,height=6cm]{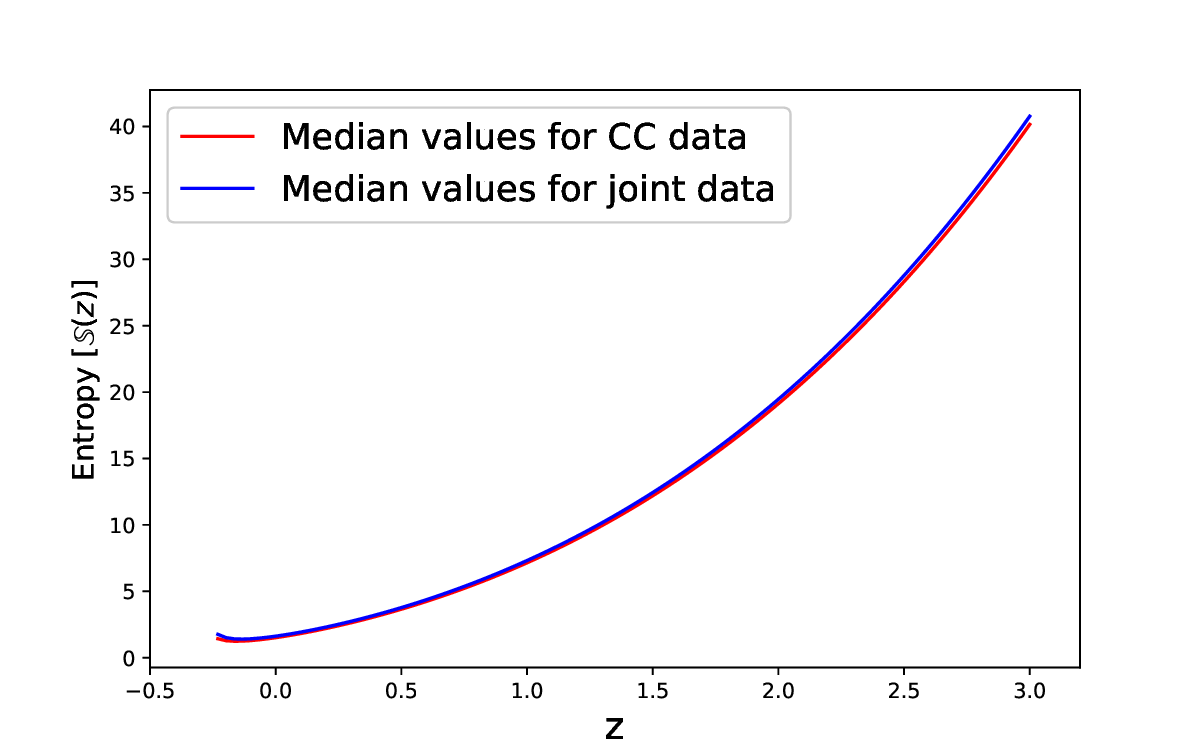}
		\caption{\textbf{For Model-I:} Plot of entropy density with $\mathit{z}$.}
		\label{fig:24}
	\end{minipage}
\end{figure}
\begin{figure}[!htb]
	\captionsetup{skip=0.4\baselineskip,size=footnotesize}
	\begin{minipage}{0.50\textwidth}
		\centering
		\includegraphics[width=7.6cm,height=6cm]{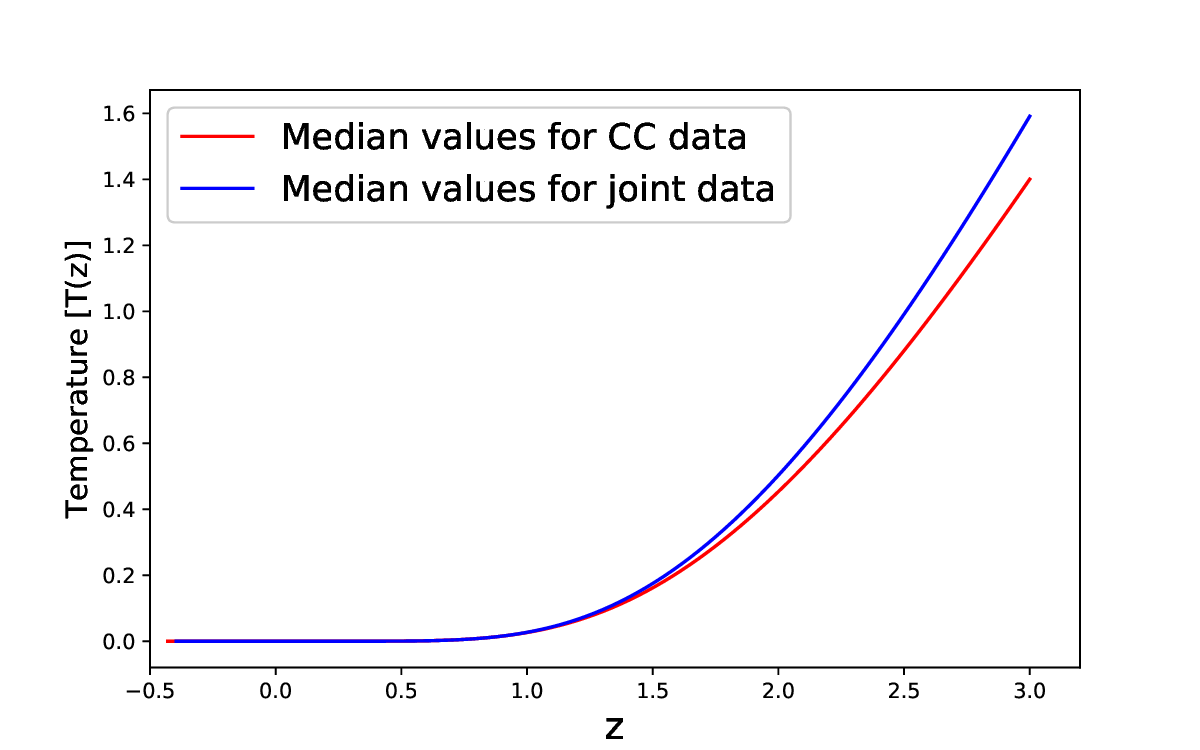}
		\caption{\textbf{For Model-II:} Plot of temperature with $\mathit{z}$.}
		\label{fig:25}
	\end{minipage}\hfill
	\begin{minipage}{0.50\textwidth}
		\centering
		\includegraphics[width=7.6cm,height=6cm]{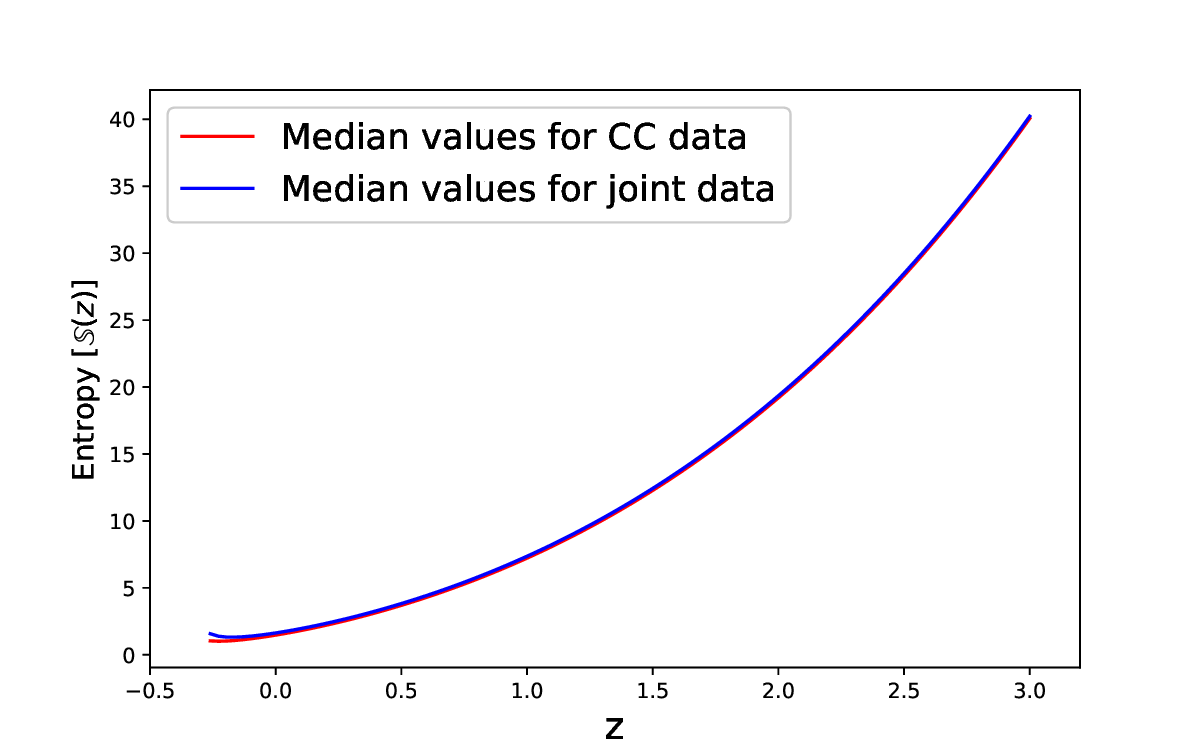}
		\caption{\textbf{For Model-II:} Plot of entropy density with $\mathit{z}$.}
		\label{fig:26}
	\end{minipage}
\end{figure}
\begin{figure}[!htb]
	\captionsetup{skip=0.4\baselineskip,size=footnotesize}
	\begin{minipage}{0.50\textwidth}
		\centering
		\includegraphics[width=7.6cm,height=6cm]{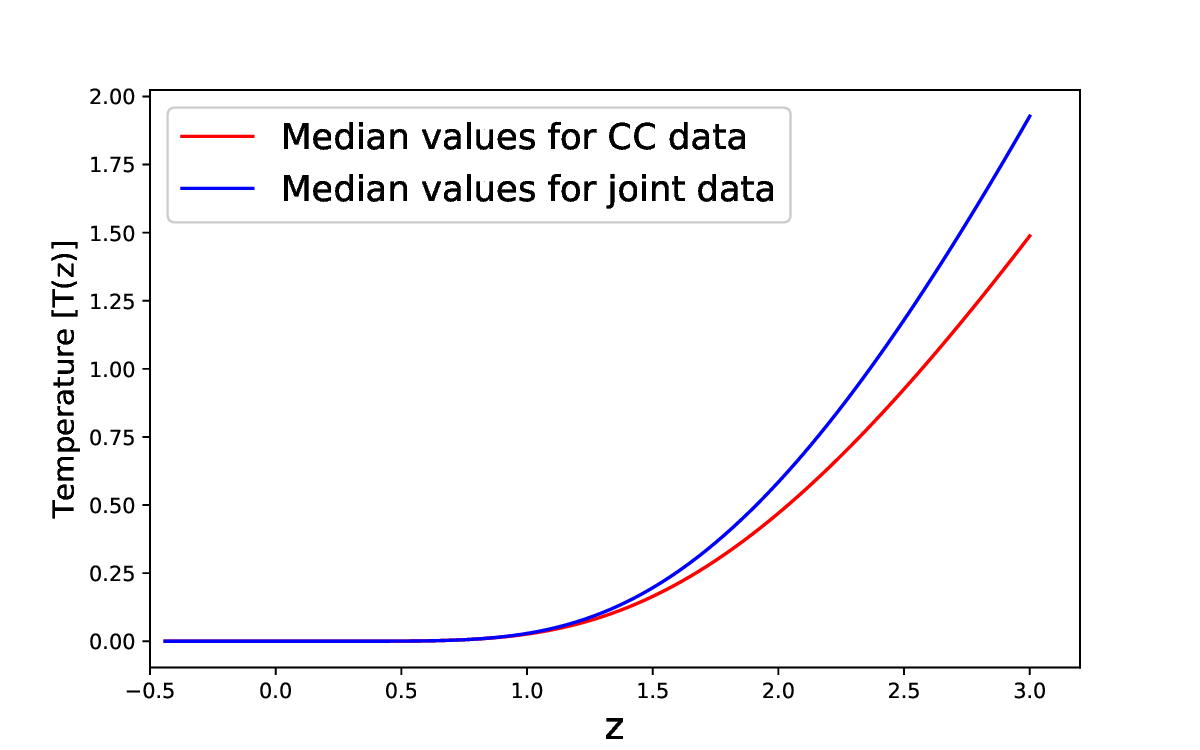}
		\caption{\textbf{For Model-III:} Plot of temperature with $\mathit{z}$.}
		\label{fig:27}
	\end{minipage}\hfill
	\begin{minipage}{0.50\textwidth}
		\centering
		\includegraphics[width=7.6cm,height=6cm]{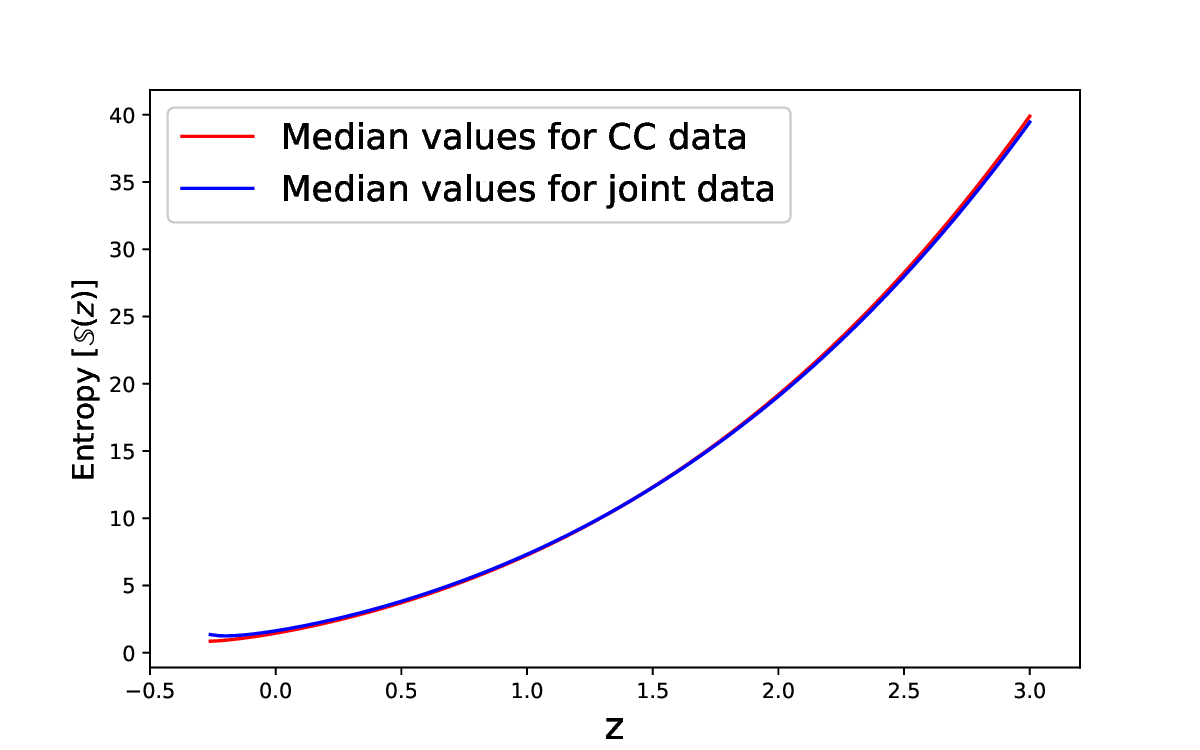}
		\caption{\textbf{For Model-III:} Plot of entropy density with $\mathit{z}$.}
		\label{fig:28}
	\end{minipage}
\end{figure}
\subsection{Thermodynamic evolution of the universe and entropy analysis}\label{sec:6.5}
To study the thermodynamic aspects of the universe in $f(R, L_m)$ gravity, we employ the laws of thermodynamics, starting with the first law for an ideal fluid enclosed in a volume $V$~\cite{Akbar2006er,bamba2012reconstruction}.
\begin{equation} {\label{48}}
	T ds = d(V \rho)+p dV,
\end{equation}
we can rewrite it as follows:
\begin{equation} {\label{49}}
	T ds = d(V(\rho+p))- V dp,
\end{equation}
and using the thermodynamic relation of the form:
\begin{equation} {\label{50}}
	dp = \left[\frac{\rho+p}{T}\right] dT.
\end{equation}
We obtain the differential form of the entropy ($s$) as:
\begin{equation} {\label{51}}
	ds = \frac{ d(V(\rho+p))}{T}- V(\rho+p) \frac{dT}{T^{2}},
\end{equation}
which can be simplified to:
\begin{equation} {\label{52}}
	ds = d\left(\frac{ V(\rho+p)}{T}\right).
\end{equation}
The integration of the above equation yields the total entropy in the form:
\begin{equation} {\label{53}}
	s = \frac{ V(\rho+p)}{T}.
\end{equation}
The definition of the entropy density ($\mathbb{S}$) is given by:
\begin{equation} {\label{54}}
	\mathbb{S}=\frac{s}{V}= \frac{ (\rho+p)}{T}= \frac{ \rho(\omega+1)}{T}.
\end{equation}
Considering a barotropic fluid described by the equation of state $p=\omega \rho$, with $0<\omega<1$, the first law of thermodynamics may accordingly be expressed as:
\begin{equation} {\label{55}}
	d(V \rho)+\omega \rho dV= (\omega+1)T d\left(\frac{V\rho}{T}\right).
\end{equation}
Upon integrating the differential relation:
\begin{equation} {\label{56}}
	\omega d\rho = (\omega+1)\rho \frac{dT}{T},
\end{equation}
which yields the temperature in terms of the energy density:
\begin{equation} {\label{57}}
	T=\rho^{\left(\frac{\omega}{\omega+1}\right)}.
\end{equation}
By inserting this relation back into the expression, we derive the entropy density in the form:
\begin{equation} {\label{58}}
	\mathbb{S}=(\omega+1)\rho^{\left(\frac{1}{\omega+1}\right)}.
\end{equation}
Using Eqs.~(\ref{36}), (\ref{38}), (\ref{40}) and (\ref{57}) together with Eqs.~(\ref{42}) to (\ref{44}) the expression for the temperature is obtained as:
\begin{equation} {\label{59}}
\resizebox{1.0\textwidth}{!}{$	
T=\left(\left(\frac{3}{2\lambda-1}\right)^{1/\lambda} \times \left\{H_{0}^{2}(1+\eta^{3\zeta})^{\frac{-8}{3}} (1+z)^{-6\zeta} \left[1+(\eta (1+z))^{3\zeta} \right]^{\frac{8}{3}}\right\}^{1/\lambda} \right)^{\left\{\frac{\left[\left(\frac{1}{3\lambda}\right) \times \left(\frac{\zeta(2-4\lambda)\left(3-\left(\eta(1+z)\right)^{3\zeta}\right)}{1+\left(\eta(1+z)\right)^{3\zeta}}\right)-1\right]}{\left(\frac{1}{3\lambda}\right) \times \left(\frac{\zeta(2-4\lambda)\left(3-\left(\eta(1+z)\right)^{3\zeta}\right)}{1+\left(\eta(1+z)\right)^{3\zeta}}\right)}\right\}}                 \text{(for Model-I)}
$}
\end{equation} 
\begin{equation} {\label{60}}
\resizebox{1.0\textwidth}{!}{$	
T=\left(\left(\frac{3}{2\lambda-1}\right)^{1/\lambda} \times \left\{H_{0}^{2}(1+\eta^{4\zeta})^{\frac{-5}{2}} (1+z)^{-8\zeta} \left[1+(\eta (1+z))^{4\zeta} \right]^{\frac{5}{2}}\right\}^{1/\lambda} \right)^{\left\{\frac{\left[\left(\frac{1}{3\lambda}\right) \times \left(\frac{\zeta(2-4\lambda)\left(4-\left(\eta(1+z)\right)^{4\zeta}\right)}{1+\left(\eta(1+z)\right)^{4\zeta}}\right)-1\right]}{\left(\frac{1}{3\lambda}\right) \times \left(\frac{\zeta(2-4\lambda)\left(4-\left(\eta(1+z)\right)^{4\zeta}\right)}{1+\left(\eta(1+z)\right)^{4\zeta}}\right)}\right\}}                 \text{(for Model-II)}
$}
\end{equation}
\begin{equation} {\label{61}}
\resizebox{1.0\textwidth}{!}{$	
T=\left(\left(\frac{3}{2\lambda-1}\right)^{1/\lambda} \times \left\{H_{0}^{2}(1+\eta^{5\zeta})^{\frac{-12}{5}} (1+z)^{-10\zeta} \left[1+(\eta (1+z))^{5\zeta} \right]^{\frac{12}{5}}\right\}^{1/\lambda} \right)^{\left\{\frac{\left[\left(\frac{1}{3\lambda}\right) \times \left(\frac{\zeta(2-4\lambda)\left(5-\left(\eta(1+z)\right)^{5\zeta}\right)}{1+\left(\eta(1+z)\right)^{5\zeta}}\right)-1\right]}{\left(\frac{1}{3\lambda}\right) \times \left(\frac{\zeta(2-4\lambda)\left(5-\left(\eta(1+z)\right)^{5\zeta}\right)}{1+\left(\eta(1+z)\right)^{5\zeta}}\right)}\right\}}                 \text{(for Model-III)}
$}
\end{equation}  
Further, using Eqs.~(\ref{36}), (\ref{38}), (\ref{40}) and (\ref{58}) together with Eqs.~(\ref{42}) to (\ref{44}) the expression for the entropy density is obtained as follows:
\begin{equation} {\label{62}}
\resizebox{1.0\textwidth}{!}{$	
\mathbb{S}= \left[\left(\frac{1}{3\lambda}\right) \times \left(\frac{\zeta(2-4\lambda)\left(3-\left(\eta(1+z)\right)^{3\zeta}\right)}{1+\left(\eta(1+z)\right)^{3\zeta}}\right)\right] \times \left(\left(\frac{3}{2\lambda-1}\right)^{1/\lambda} \times \left\{H_{0}^{2}(1+\eta^{3\zeta})^{\frac{-8}{3}} (1+z)^{-6\zeta} \left[1+(\eta (1+z))^{3\zeta} \right]^{\frac{8}{3}}\right\}^{1/\lambda} \right)^{\left(\frac{1}{\left(\frac{1}{3\lambda}\right) \times \left(\frac{\zeta(2-4\lambda)\left(3-\left(\eta(1+z)\right)^{3\zeta}\right)}{1+\left(\eta(1+z)\right)^{3\zeta}}\right)}\right)}                 \text{(for Model-I)}
$}
\end{equation} 
\begin{equation} {\label{63}}
\resizebox{1.0\textwidth}{!}{$	
\mathbb{S}=\left[\left(\frac{1}{3\lambda}\right) \times \left(\frac{\zeta(2-4\lambda)\left(4-\left(\eta(1+z)\right)^{4\zeta}\right)}{1+\left(\eta(1+z)\right)^{4\zeta}}\right)\right] \times \left(\left(\frac{3}{2\lambda-1}\right)^{1/\lambda} \times \left\{H_{0}^{2}(1+\eta^{4\zeta})^{\frac{-5}{2}} (1+z)^{-8\zeta} \left[1+(\eta (1+z))^{4\zeta} \right]^{\frac{5}{2}}\right\}^{1/\lambda} \right)^{\left(\frac{1}{\left(\frac{1}{3\lambda}\right) \times \left(\frac{\zeta(2-4\lambda)\left(4-\left(\eta(1+z)\right)^{4\zeta}\right)}{1+\left(\eta(1+z)\right)^{4\zeta}}\right)}\right)}                 \text{(for Model-II)}
$}
\end{equation}
\begin{equation} {\label{64}}
\resizebox{1.0\textwidth}{!}{$	
\mathbb{S}= \left[\left(\frac{1}{3\lambda}\right) \times \left(\frac{\zeta(2-4\lambda)\left(5-\left(\eta(1+z)\right)^{5\zeta}\right)}{1+\left(\eta(1+z)\right)^{5\zeta}}\right)\right] \times \left(\left(\frac{3}{2\lambda-1}\right)^{1/\lambda} \times \left\{H_{0}^{2}(1+\eta^{5\zeta})^{\frac{-12}{5}} (1+z)^{-10\zeta} \left[1+(\eta (1+z))^{5\zeta} \right]^{\frac{12}{5}}\right\}^{1/\lambda} \right)^{\left(\frac{1}{\left(\frac{1}{3\lambda}\right) \times \left(\frac{\zeta(2-4\lambda)\left(5-\left(\eta(1+z)\right)^{5\zeta}\right)}{1+\left(\eta(1+z)\right)^{5\zeta}}\right)}\right)}                 \text{(for Model-III)}
$}
\end{equation}  
The evolutionary profiles of temperature and entropy density for the proposed cosmological models, as shown in Figs.~(\ref{fig:23}) to (\ref{fig:28}), exhibits a clear dependence on redshift under observational constraints. From Figs.~(\ref{fig:23}), (\ref{fig:25}) and (\ref{fig:27}), suggests that the universe remained extremely hot during its initial stages, with temperature rising toward higher redshift values. This behavior is consistent with standard cosmological expectations and confirms that the models can effectively reproduce the expected thermal history of the universe. In the far-future regime ($z<0$), the temperature approaches an asymptotic constant value, indicating a tendency toward thermal stabilization in the low-energy regime. Furthermore, the behavior of the entropy density illustrated in Figs.~(\ref{fig:24}), (\ref{fig:26}) and (\ref{fig:28}), is obtained using the dimensionless energy density normalized by the critical density scale $3H_0^2$. The findings suggest an increase in entropy density with redshift, implying higher entropy levels at earlier cosmic epochs. This trend is consistent with the conventional cosmological picture in which energy density and temperature are larger at early times. As the universe expands, the entropy density decreases due to volume expansion and energy dilution; however, the total entropy is expected to increase, adhering to the generalized second law. Moreover, the relation $\mathbb{S} \propto \rho^{\left(\frac{1}{\omega+1}\right)}$ signifies the interaction between thermodynamic quantities and the underlying energy density dynamics. Collectively, these results demonstrate the robustness of the thermodynamic analysis and reinforce the suitability of $f(R, L_{m})$ gravity as a viable framework for describing the evolution of the universe.
\subsection{Estimation of the age of the universe}\label{sec:6.6}
The redshift dependence of the cosmic age function $t(z)$ for the present cosmological framework is examined in~\cite{tong2009cosmic}:
\begin{equation} {\label{47}}
	t(z) = \int_{z}^{\infty} \frac{dz}{(1+z)H(z)}.
\end{equation}
The age of the universe at the present epoch, denoted by $t_{0}$, is evaluated numerically from the integral involving the Hubble parameter $H(z)$ corresponding to Eqs.~(\ref{23}) to (\ref{25}) at $z=0$. Using the CC dataset, Model-I predicts $t_{0}=13.20$ Gyr, while the joint dataset gives $t_{0}=13.08$ Gyr. For Model-II, the estimated present ages are $t_{0}=13.20$ Gyr and $t_{0}=13.05$ Gyr for the CC and combined datasets, respectively. Similarly, Model-III provides $t_{0}=13.18$ Gyr from the CC analysis and $t_{0}=13.02$ Gyr from the joint analysis. Notably, these estimates are in close agreement with the age of the universe inferred from the $\Lambda$CDM cosmology based on the Planck observations~\cite{2020A&A...641A...6P}.
\section{Conclusions}\label{sec:7}
In this paper, we study and probe the cosmic evolution of a spatially flat FLRW universe within the framework of $f(R, L_{m})$ gravity by adopting a well-motivated functional form of the theory. We analyze the cosmological dynamics by considering a specific parametrized form of the Hubble parameter $H(t)$, which gives rise to three distinct cosmological models, namely Model-I, Model-II and Model-III (as summarized in Table~(\ref{table:1})). These models provide a flexible and dynamically enriched description of the cosmic expansion history, enabling a unified representation of different evolutionary phases of the universe. The model parameters have been constrained using the cosmic chronometer (CC) dataset as well as the joint (CC+Pantheon) dataset through a Bayesian MCMC analysis and the corresponding results are presented in Tables~(\ref{table:2}) to (\ref{table:4}). The best-fit curves of the reconstructed Hubble parameter $H(z)$, illustrated in Fig.~(\ref{fig:1}), demonstrate the compatibility of all three models with the measured cosmic chronometer observations.
\vspace{0.2cm}\\
The evolution of the deceleration parameter (Figs.~(\ref{fig:5}) to (\ref{fig:7})), clearly shows that all three models successfully capture the transition from an early decelerated epoch to the current phase of accelerated expansion of the universe. The present-day values are obtained as $q_{0} = -0.5323$ (CC) and $q_{0} = -0.5338$ (joint) for Model-I, $q_{0} = -0.5108$ (CC) and $q_{0} = -0.5254$ (joint) for Model-II, and $q_{0} = -0.4995$ (CC) and $q_{0} = -0.5238$ (joint) for Model-III, thereby firmly supporting the current accelerated state of cosmic expansion. The transition redshift is found to be $z_{t} = 0.6547$ and $z_{t} = 0.6523$ for Model-I, $z_{t} = 0.6442$ and $z_{t} = 0.6495$ for Model-II, and $z_{t} = 0.6486$ and $z_{t} = 0.6504$ for Model-III (for CC and joint datasets, respectively), which are fully consistent with observational expectations. An important feature emerging from the present analysis is that all three models naturally permit a transition into a super-accelerated (phantom) regime at late times. In particular, the models exhibit a smooth evolution from a matter-dominated phase at high redshift to a dark energy-dominated accelerated phase at low redshift, thereby providing a flexible and physically consistent description of cosmic expansion.
\vspace{0.2cm}\\
The physical viability of the models is further reinforced by the behavior of the energy density and pressure, as shown in Figs.~(\ref{fig:8}) to (\ref{fig:13}). The energy density remains strictly positive throughout the entire cosmic evolution, ensuring physical admissibility, while the pressure evolves from positive values at early times to negative values at late times. This negative pressure component plays a crucial role in driving the accelerated expansion of the universe. The corresponding evolution of the EoS parameter (Figs.~(\ref{fig:14}) to (\ref{fig:16})) indicates that the present universe resides within the quintessence regime ($-1 < \omega < -\tfrac{1}{3}$). The present-day values of the EoS parameter are obtained as $\omega_{0} = -0.7015$ (CC) and $\omega_{0} = -0.7022$ (joint) for Model-I, $\omega_{0} = -0.6875$ (CC) and $\omega_{0} = -0.6968$ (joint) for Model-II, and $\omega_{0} = -0.6802$ (CC) and $\omega_{0} = -0.6957$ (joint) for Model-III. Furthermore, all models exhibit a gradual transition toward the phantom regime at late times, suggesting a quintom-like dynamical nature of dark energy. At sufficiently high redshifts, the models asymptotically approach a matter-dominated phase, in accordance with the requirements of structure formation.
\vspace{0.2cm}\\
The analysis of energy conditions (in Figs.~(\ref{fig:17}) to (\ref{fig:19})) reveals that the Null, Weak and Dominant Energy Conditions are satisfied up to the present epoch, thereby ensuring the physical consistency of the models within the observed regime. In contrast, the SEC is violated, which is a necessary condition for explaining the observed accelerated expansion. At late times, the violation of the NEC further signals the emergence of phantom behavior. Additional insight is provided by geometrical diagnostics such as the statefinder $\{r, s\}$ plane (in Figs.~(\ref{fig:20}) to (\ref{fig:22})), where all models evolve from the Chaplygin gas regime, pass through the $\Lambda$CDM fixed point and asymptotically approach a unified dark sector scenario. This behavior highlights the rich dynamical structure of the proposed framework.
\vspace{0.2cm}\\
From a thermodynamic perspective, the models are examined through the evolution profiles of temperature and entropy density (in Figs.~(\ref{fig:23}) to (\ref{fig:28})). An increasing trend of $T(z)$ with redshift indicates that the early universe was hotter, followed by progressive cooling as expansion proceeds. Likewise, $\mathbb{S}(z)$ attains higher values in the early universe, consistent with greater energy content and shows a gradual decline as the universe expands. This behavior reflects a thermodynamically consistent evolution within the framework of the study. The estimated age of the universe is found to be $t_{0} = 13.20$ Gyr and $t_{0} = 13.08$ Gyr for Model-I, $t_{0} = 13.20$ Gyr and $t_{0} = 13.05$ Gyr for Model-II, and $t_{0} = 13.18$ Gyr and $t_{0} = 13.02$ Gyr for Model-III, corresponding to the CC and joint datasets, respectively. These values are in close agreement with current observational estimates, further supporting the viability of our models.
\vspace{0.2cm}\\
In summary, the present study provides a comprehensive analysis of cosmic evolution within the $f(R, L_m)$ gravity framework using parametrized forms of the Hubble parameter leading to three newly constructed cosmological models. All three models successfully reproduce the key features of the universe, including the transition from deceleration to acceleration, the dominance of dark energy at late times and the possibility of phantom behavior in the future. The consistency with observational data, together with well-behaved physical, geometrical and thermodynamic properties, demonstrates that the proposed models offer a viable and robust description of late-time cosmic dynamics. Therefore, the present models may serve as reliable candidates for investigating the late-time dynamics of the universe within the framework of modified gravity.
\section*{\textbf{Acknowledgements}}
GPS acknowledges the Inter-University Centre for Astronomy and Astrophysics (IUCAA), Pune, India for the support provided through the Visiting Associateship Programme.


\begin{thebibliography}{0}
	\bibitem{1998AJ....116.1009R} A. G. Riess, A. V.  Filippenko, P. Challis, A. Clocchiatti, A. Diercks  et al., Observational Evidence from Supernovae for an Accelerating Universe and a Cosmological Constant, Astronomical Journal, \textbf{116}, 1009-1038 (1998) \url{https://doi.org/10.1086/300499} 
	\bibitem{1999ApJ...517..565P}  S. Perlmutter, G. Aldering, G. Goldhaber, R. A. Knop, P. Nugent et al., Measurements of $\Omega$ and $\Lambda$ from 42 High-Redshift Supernovae, Astrophysical Journal, \textbf{517}, 565–586 (1999)  \url{https://doi.org/10.1086/307221} 
	\bibitem{2020A&A...641A...6P} N. Aghanim, Y. Akrami, M. Ashdown, J. Aumont, C. Baccigalupi, et al., Planck 2018 results. VI. Cosmological parameters, Astronomy and Astrophysics, \textbf{641}, A6 (2020) \url{https://doi.org/10.1051/0004-6361/201833910}
	\bibitem{weinberg1989cosmological} S. Weinberg, The cosmological constant problem, Reviews of Modern Physics, \textbf{61}, 1 (1989) \url{https://doi.org/10.1103/RevModPhys.61.1}
	\bibitem{di2021realm} E. Di Valentino, O. Mena, S. Pan, L. Visinelli, W. Yang, et al., In the realm of the Hubble tension—a review of solutions, Classical and Quantum Gravity, \textbf{38}, 153001 (2021) \url{https://doi.org/10.1088/1361-6382/ac086d}
	\bibitem{carroll2001cosmological} S. M. Carroll, The cosmological constant, Living Reviews in Relativity, \textbf{4}, 1-56 (2001) \url{https://doi.org/10.12942/lrr-2001-1}
	\bibitem{Kerner} R. Kerner, Cosmology without singularity and nonlinear gravitational Lagrangians, General Relativity and Gravitation, \textbf{14}, 453–469 (1982) \url{https://doi.org/10.1007/BF00756329}
	\bibitem{buchdahl1970non} H. A. Buchdahl, Non-linear Lagrangians and cosmological theory, Monthly Notices of the Royal Astronomical Society, \textbf{150}, 1-8 (1970) \url{https://doi.org/10.1093/mnras/150.1.1}
	\bibitem{nojiri2011unified} S. Nojiri, S. D. Odintsov, Unified cosmic history in modified gravity: from $f(R)$ theory to Lorentz non-invariant models, Physics Reports, \textbf{505}, 59-144 (2011) \url{https://doi.org/10.1016/j.physrep.2011.04.001}
	\bibitem{nojiri2017modified} S. Nojiri, S. D. Odintsov, V. K. Oikonomou, Modified gravity theories on a nutshell: Inflation, bounce and late-time evolution, Physics Reports, \textbf{692}, 1-104 (2017) \url{https://doi.org/10.1016/j.physrep.2017.06.001}
	\bibitem{harko2011f} T. Harko, F. S. N. Lobo, S. Nojiri, S. D. Odintsov, $f(R,T)$ gravity, Physical Review D, \textbf{84}, 024020 (2011) \url{https://doi.org/10.1103/PhysRevD.84.024020}
	\bibitem{cai2016f} Yi-Fu Cai, S. Capozziello, M. De Laurentis, E. N. Saridakis, $f(T)$ teleparallel gravity and cosmology, Reports on Progress in Physics, \textbf{79}, 106901 (2016) \url{https://doi.org/10.1088/0034-4885/79/10/106901}
	\bibitem{capozziello2011cosmography} S. Capozziello, V. F. Cardone, H. Farajollahi, A. Ravanpak, Cosmography in $f(T)$ gravity, Physical Review D, \textbf{84}, 043527 (2011) \url{https://doi.org/10.1103/PhysRevD.84.043527}
	\bibitem{capozziello2019extended} S. Capozziello, R. D'Agostino, O. Luongo, Extended gravity cosmography, International Journal of Modern Physics D, \textbf{28}, 1930016 (2019) \url{https://doi.org/10.1142/S0218271819300167}
	\bibitem{bamba2010finite} K. Bamba, S. D. Odintsov, L. Sebastiani, S. Zerbini, Finite-time future singularities in modified Gauss-Bonnet and $f(R,G)$ gravity and singularity avoidance, The European Physical Journal C, \textbf{67}, 295-310 (2010) \url{https://doi.org/10.1140/epjc/s10052-010-1292-8}
	\bibitem{capozziello2023role} S. Capozziello, V. De Falco, C. Ferrara, The role of the boundary term in $f(Q, B)$ symmetric teleparallel gravity,The European Physical Journal C, \textbf{83}, 915 (2023) \url{https://doi.org/10.1140/epjc/s10052-023-12072-y}
	\bibitem{lalke2023late} A. R. Lalke, G. P. Singh, A. Singh, Late-time acceleration from ekpyrotic bounce in $f(Q, T)$ gravity, International Journal of Geometric Methods in Modern Physics, \textbf{20}, 2350131 (2023) \url{https://doi.org/10.1142/S0219887823501311}
	\bibitem{kotambkar2017anisotropic} S. Kotambkar, G. P. Singh, R. Kelkar, B. K. Bishi, Anisotropic Bianchi type I cosmological models with generalized Chaplygin gas and dynamical gravitational and cosmological constants, Communications in Theoretical Physics, \textbf{67}, 222 (2017) \url{https://doi.org/10.1088/0253-6102/67/2/222}
	\bibitem{singh2002viscous} G. P. Singh, R. V. Deshpande, T. Singh, Viscous cosmological models with particle creation in Brans-Dicke theory, Astrophysics and space science, \textbf{282}, 489--498 (2002) \url{https://doi.org/10.1023/A:1020963219962}
	\bibitem{goswami2024flrw} G. K. Goswami, R. Rani, J. K. Singh, A. Pradhan, FLRW cosmology in Weyl type $f(Q)$ gravity and observational constraints, Journal of High Energy Astrophysics, \textbf{43}, 105--113 (2024) \url{https://doi.org/10.1016/j.jheap.2024.06.011}
	\bibitem{patle2026dynamical} K. R. Patle, G. P. Singh, R. Garg, Dynamical constraints on variable vacuum energy in Brans-Dicke theory, arXiv preprint arXiv:2601.00419, (2026) \url{https://doi.org/10.48550/arXiv.2601.00419}
	\bibitem{varela2025cosmological} M. B. Varela, O. Bertolami, Is cosmological data suggesting a nonminimal coupling between matter and gravity?, Physics of the Dark Universe, \textbf{48}, 101861 (2025) \url{https://doi.org/10.1016/j.dark.2025.101861}
	\bibitem{singh2024conservative} K. N. Singh, G. R. P. Teruel, S. K. Maurya, T. Chowdhury, F. Rahaman, Conservative wormholes in generalized $K(R, T)$ function, Journal of High Energy Astrophysics, \textbf{44}, 132--145 (2024) \url{https://doi.org/10.1016/j.jheap.2024.09.009}
	\bibitem{singh2022cosmic} A. Singh, G. P. Singh, A. Pradhan, Cosmic dynamics and qualitative study of Rastall model with spatial curvature, International Journal of Modern Physics A, \textbf{37}, 2250104 (2022) \url{https://doi.org/10.1142/S0217751X22501044}
	\bibitem{hulke2020variable} N. Hulke, G. P. Singh, B. K. Bishi, A. Singh, Variable Chaplygin gas cosmologies in $f(R, T)$ gravity with particle creation, New Astronomy, \textbf{77}, 101357 (2020) \url{https://doi.org/10.1016/j.newast.2020.101357}
	\bibitem{singh505abc} G. P. Singh, R. Garg, A. Singh, A generalized $\Lambda$CDM model with parameterized Hubble parameter in particle creation, viscous and $f(R)$ model framework, International Journal of Geometric Methods in Modern Physics, 2550111 (2025) \url{https://doi.org/10.1142/S0219887825501117}
	\bibitem{shukla2025multi} B. K. Shukla, S. Sahlu, D. Sofuo{\u{g}}lu, P. Mishra, A. H. Alfedeel, Multi-components fluid in $f(R, T)$ gravity with observational constraints, The European Physical Journal Plus, \textbf{140}, 1--14 (2025)
	\url{https://doi.org/10.1140/epjp/s13360-025-06200-8}
	\bibitem{harko2010f} T. Harko, F. S. N. Lobo, $f(R, L_m)$ gravity, The European Physical Journal C, \textbf{70}, 373--379 (2010) \url{https://doi.org/10.1140/epjc/s10052-010-1467-3}
	\bibitem{Faraoni2004pi} V. Faraoni, Cosmology in scalar tensor gravity, Springer (2004) \url{https://doi.org/10.1007/978-1-4020-1989-0}
	\bibitem{zhang2007behavior} P. Zhang, Behaviour of $f(R)$ gravity in the solar system, galaxies, and clusters, Physical Review D, \textbf{76}, 024007 (2007) \url{https://doi.org/10.1103/PhysRevD.76.024007}
	\bibitem{bertolami2008general} O. Bertolami, J. Páramos, S. G. Turyshev, General Theory of Relativity: Will it survive the next decade?, in: Lasers, Clocks and Drag-Free Control: Exploration of Relativistic Gravity in Space, Springer, 27--74 (2008) \url{https://doi.org/10.48550/arXiv.gr-qc/0602016}
	\bibitem{Rahaman2009solar} F. Rahaman, S. Ray, M. Kalam, M. Sarker, Do Solar system tests permit higher dimensional general relativity?, International Journal of Theoretical Physics, \textbf{48}, 3124--3138 (2009) \url{https://doi.org/10.1007/s10773-009-0110-2}
	\bibitem{nojiri2004gravity} S. Nojiri, S. D. Odintsov, Gravity assisted dark energy dominance and cosmic acceleration, Physics Letters B, \textbf{599}, 137--142 (2004) \url{https://doi.org/10.1016/j.physletb.2004.08.045}
	\bibitem{allemandi2005dark} G. Allemandi, A. Borowiec, M. Francaviglia, S. D. Odintsov, Dark energy dominance and cosmic acceleration in first-order formalism, Physical Review D, \textbf{72}, 063505 (2005) \url{https://doi.org/10.1103/PhysRevD.72.063505}
	\bibitem{manna2023gravity} G. Manna, A. Panda, A. Karmakar, S. Ray, M. R. Islam, $f(R, L_x)$-gravity in the context of dark energy with power law expansion and energy conditions, Chinese Physics C, \textbf{47}, 025101 (2023) \url{https://doi.org/10.1088/1674-1137/ac9fbe}
	\bibitem{lobo2015extended} F. S. N. Lobo, T. Harko, Extended $f(R, L_m)$ theories of gravity, THE THIRTEENTH MARCEL GROSSMANN MEETING, 1164--1166 (2015) \url{https://doi.org/10.48550/arXiv.1211.0426}
	\bibitem{jaybhaye2022cosmology} L. V. Jaybhaye, R. Solanki, S. Mandal, P. K. Sahoo, Cosmology in $f(R, L_m)$ gravity, Physics Letters B, \textbf{831}, 137148 (2022) \url{https://doi.org/10.1016/j.physletb.2022.137148}
	\bibitem{doi:10.1142/S0219887823501050} A. Pradhan, D. C. Maurya, G. K. Goswami, A. Beesham, Modeling transit dark energy in $f(R, L_m)$ gravity, International Journal of Geometric Methods in Modern Physics, \textbf{20}, 2350105 (2023) \url{https://doi.org/10.1142/S0219887823501050}
	\bibitem{koussour2024bouncing} M. Koussour, N. Myrzakulov, J. Rayimbaev, A. H. A. Alfedeel, H. M. Elkhair, Bouncing behavior in $f(R, L_m)$ gravity: Phantom crossing and energy conditions, International Journal of Geometric Methods in Modern Physics, \textbf{21}, 2450184 (2024) \url{https://doi.org/10.1142/S0219887824501846}
	\bibitem{mustafa2024dynamical} G. Mustafa, F. Javed, S. K. Maurya, M. Govender, A. Saleem, Dynamical stability of new wormhole solutions via cold dark matter and solitonic quantum wave halos in $f(R, L_m)$ gravity, Physics of the Dark Universe, \textbf{45}, 101508 (2024)
	\url{https://doi.org/10.1016/j.dark.2024.101508}
	\bibitem{myrzakulova2024investigating} S. Myrzakulova, M. Koussour, N. Myrzakulov, Investigating the dark energy phenomenon in $f(R, L_m)$ cosmological models with observational constraints, Physics of the Dark Universe, \textbf{43}, 101399 (2024) \url{https://doi.org/10.1016/j.dark.2023.101399}
	\bibitem{kavya2022constraining} N. S. Kavya, V. Venkatesha, S. Mandal, P. K. Sahoo, Constraining anisotropic cosmological model in $f(R, L_m)$ Gravity, Physics of the Dark Universe, \textbf{38}, 101126 (2022) \url{https://doi.org/10.1016/j.dark.2022.101126}
	\bibitem{devi2024constraining} Y. K. Devi, S. Narawade, B. Mishra, Constraining parameters for the accelerating universe in $f(R, L_m)$ gravity, Physics of the Dark Universe, \textbf{46}, 101640 (2024) \url{https://doi.org/10.1016/j.dark.2024.101640}
	\bibitem{zeyauddin2024anisotropic} M. Zeyauddin, A. Dixit, A. Pradhan, Anisotropic dark energy models in $f(R, L_m)$-gravity, International Journal of Geometric Methods in Modern Physics, \textbf{21}, 2450038 (2024) \url{https://doi.org/10.1142/S0219887824500385}
	\bibitem{maurya2024bianchi} D. C. Maurya, Bianchi-I dark energy cosmological model in $f(R, L_m)$-gravity, International Journal of Geometric Methods in Modern Physics, \textbf{21}, 2450072 (2024) \url{https://doi.org/10.1142/S0219887824500725}
	\bibitem{sahlu2024cosmology} S. Sahlu, A. H. Alfedeel, A. Abebe, The cosmology of $f(R, L_m)$ gravity: constraining the background and perturbed dynamics, The European Physical Journal C, \textbf{84}, 982 (2024) \url{https://doi.org/10.1140/epjc/s10052-024-13307-2}
	\bibitem{bhardwaj2025cosmological} V. K. Bhardwaj, S. Ray, Cosmological model in the framework of $f(R, L_m)$ gravity with quadratic equation of state parameter, Physics of the Dark Universe, 101930 (2025) \url{https://doi.org/10.1016/j.dark.2025.101930}
	\bibitem{jaybhaye2022constraints} L. V. Jaybhaye, S. Mandal, P. K. Sahoo, Constraints on energy conditions in $f(R, L_m)$ gravity, International Journal of Geometric Methods in Modern Physics, \textbf{19}, 2250050 (2022) \url{https://doi.org/10.1142/S0219887822500505}
	\bibitem{shukla2025dynamical} A. Shukla, R. Raushan, R. Chaubey, Dynamical Systems Analysis of $f(R, L_m)$ Gravity Model, International Journal of Geometric Methods in Modern Physics, 2550118 (2025) \url{https://doi.org/10.1142/S021988782550118X}
	\bibitem{myrzakulov2024linear} Y. Myrzakulov, O. Donmez, G. D. A. Yildiz, E. Gudekli, S. Muminov et al., Linear redshift parametrization of deceleration parameter in $f(R, L_m)$ gravity, Physics of the Dark Universe, \textbf{45}, 101545 (2024) \url{https://doi.org/10.1016/j.dark.2024.101545}
	\bibitem{singh2024consequence} J. K. Singh, A. Singh, H. Balhara, J. R. L. Santos, The consequence of higher-order curvature-based constraints on $f(R, L_m)$ gravity, Annals of Physics, \textbf{469}, 169781 (2024) \url{https://doi.org/10.1016/j.aop.2024.169781}
	\bibitem{chaudhary2024existence} S. Chaudhary, J. Kumar, S. K. Maurya, S. Kiroriwal, A. Aziz, On the existence and stability of traversable wormhole solutions with novel shapefunctions in the framework of $f(R, L_m)$ gravity, Communications in Theoretical Physics, \textbf{76}, 055403 (2024) \url{https://doi.org/10.1088/1572-9494/ad3544}
	\bibitem{wu2014constraints} Y. B. Wu, Y. Y. Zhao, Y. Y. Jin, L. L. Lin, J. B. Lu et al., Constraints of energy conditions and DK instability criterion on $f(R, L_m)$ gravity models, Modern Physics Letters A, \textbf{29}, 1450089 (2014) \url{https://doi.org/10.1142/S0217732314500898}
	\bibitem{kavya2023static} N. S. Kavya, V. Venkatesha, G. Mustafa, P. K. Sahoo, S. V. D. Rashmi, Static traversable wormhole solutions in $f(R, L_m)$ gravity, Chinese Journal of Physics, \textbf{84}, 1--11 (2023) \url{https://doi.org/10.1016/j.cjph.2023.05.002}
	\bibitem{Shafieloo} A. Shafieloo, A. G. Kim, E. V. Linder, Model independent tests of cosmic growth versus expansion, Physical Review D, \textbf{87}, 2 (2013) \url{https://doi.org/10.1103/PhysRevD.87.023520}
	\bibitem{Amendola} A. Gómez-Valent, L. Amendola, H0 from cosmic chronometers and Type Ia supernovae, with Gaussian Processes and the novel Weighted Polynomial Regression method, Journal of Cosmology and Astroparticle Physics, \textbf{04}, 051 (2018) \url{https://doi.org/10.1088/1475-7516/2018/04/051}
	\bibitem{Koussour} M. Koussour, S. K. J. Pacif, M. Bennai, P. K. Sahoo, A New Parametrization of Hubble Parameter in $f(Q)$ Gravity, Fortschritte der Physik, \textbf{71}, 2200172 (2023) \url{https://doi.org/10.1002/prop.202200172}
	\bibitem{RitikaNagpal} J. K. Singh, R. Nagpal, FLRW cosmology with EDSFD parametrization, The European Physical Journal C, \textbf{80}, 295 (2020) \url{https://doi.org/10.1140/epjc/s10052-020-7827-8}
	\bibitem{he2024new} T. Y. He, J. J. Yin, Z. Y. Wang, Z. W. Han, R. J. Yang, A new parametrization of Hubble function and Hubble tension, Journal of Cosmology and Astroparticle Physics, \textbf{2024}, 028 (2024) \url{https://doi.org/10.1088/1475-7516/2024/09/028}
	\bibitem{RoyGoswami} N. Roy, S. Goswami, S. Das, Quintessence or phantom: Study of scalar field dark energy models through a general parametrization of the Hubble parameter, Physics of the Dark Universe, \textbf{36}, 101037 (2022) \url{https://doi.org/10.1016/j.dark.2022.101037}
	\bibitem{partridge2004introduction} B. Ryden, Introduction to Cosmology, Addison-Wesley San Francisco, (2003)
	\bibitem{harko2014generalized} T. Harko, F. S. N. Lobo, Generalised curvature-matter couplings in modified gravity, Galaxies, \textbf{2}, 410--465 (2014) \url{https://doi.org/10.3390/galaxies2030410}
	\bibitem{harko2015gravitational} T. Harko, F. S. N. Lobo, J. P. Mimoso, D. Pavón, Gravitational induced particle production through a nonminimal curvature--matter coupling, The European Physical Journal C, \textbf{75}, 1--15 (2015) \url{https://doi.org/10.1140/epjc/s10052-015-3620-5}
	\bibitem{shafieloo2013model} A. Shafieloo, A. G. Kim, E. V. Linder, Model independent tests of cosmic growth versus expansion, Physical Review D, \textbf{87}, 023520 (2013) \url{https://doi.org/10.1103/PhysRevD.87.023520}
	\bibitem{banerjee2005acceleration} N. Banerjee, S. Das, Acceleration of the universe with a simple trigonometric potential, General Relativity and Gravitation, \textbf{37}, 1695–1703 (2005) \url{https://doi.org/10.1007/s10714-005-0152-6}
	\bibitem{cunha2008transition} J. Cunha, J. A. S. d. Lima, Transition redshift: new kinematic constraints from supernovae, Monthly Notices of the Royal Astronomical Society, \textbf{390}, 210–217 (2008) \url{https://doi.org/10.1111/j.1365-2966.2008.13640.x}
	\bibitem{escamilla2022dynamical} C. Escamilla-Rivera, A. N{\'a}jera, Dynamical dark energy models in the light of gravitational-wave transient catalogues, Journal of Cosmology and Astroparticle Physics, \textbf{2022}, 060 (2022) \url{https://doi.org/10.1088/1475-7516/2022/03/060}
	\bibitem{myrzakulov2023quintessence} N. Myrzakulov, M. Koussour, A. Mussatayeva, Quintessence-like features in the late-time cosmological evolution of $f(Q)$ symmetric teleparallel gravity, Chinese Journal of Physics, \textbf{85}, 345--358 (2023) \url{https://doi.org/10.1016/j.cjph.2023.07.003}
	\bibitem{NagpalPacif} R. Nagpal, S. K. J. Pacif, F. Atamurotov, R. Pati, Dark sector interactions: Probing the Hubble parameter and the sound horizon, Annals of Physics, \textbf{483}, 170249 (2025) \url{https://doi.org/10.1016/j.aop.2025.170249}
	\bibitem{yadav2024reconstructing} A. K. Yadav, S. R. Bhoyar, M. C. Dhabe, S. H. Shekh, N. Ahmad, Reconstructing $f(Q)$ gravity from parameterization of the Hubble parameter and observational constraints, Journal of High Energy Astrophysics, \textbf{43}, 114--125 (2024) \url{https://doi.org/10.1016/j.jheap.2024.06.012}
	\bibitem{pacif2017reconstruction} S. K. J. Pacif, R. Myrzakulov, S. Myrzakul, Reconstruction of cosmic history from a simple parametrization of H, International Journal of Geometric Methods in Modern Physics, \textbf{14}, 1750111 (2017) \url{https://doi.org/10.1142/S0219887817501110}
	\bibitem{foreman2013emcee} D. Foreman-Mackey, D. W. Hogg, D. Lang, J. Goodman, emcee: the MCMC hammer, Publications of the Astronomical Society of the Pacific, \textbf{125}, 306 (2013) \url{https://doi.org/10.1086/670067}
	\bibitem{simon2005constraints} J. Simon, L. Verde, R. Jimenez, Constraints on the redshift dependence of the dark energy potential, Physical Review D, \textbf{71}, 123001 (2005) \url{https://doi.org/10.1103/PhysRevD.71.123001}
	\bibitem{sharov2018predictions} G. S. Sharov, V. O. Vasiliev, How predictions of cosmological models depend on Hubble parameter data sets, arXiv preprint arXiv:1807.07323 (2018) \url{https://doi.org/10.26456/mmg/2018-611}
	\bibitem{stern2010cosmic} D. Stern, R. Jimenez, L. Verde, M. Kamionkowski, S. A. Stanford, Cosmic chronometers: constraining the equation of state of dark energy. I: $H(z)$ measurements, Journal of Cosmology and Astroparticle Physics, \textbf{2010}, 008 (2010) \url{https://doi.org/10.1088/1475-7516/2010/02/008}
	\bibitem{moresco2015raising} M. Moresco, Raising the bar: new constraints on the Hubble parameter with cosmic chronometers at $z~ 2$, Monthly Notices of the Royal Astronomical Society: Letters, \textbf{450}, L16--L20 (2015) \url{https://doi.org/10.1093/mnrasl/slv037}
	\bibitem{jimenez2002constraining} R. Jimenez, A. Loeb, Constraining cosmological parameters based on relative galaxy ages, The Astrophysical
	Journal, \textbf{573}, 37 (2002) \url{https://doi.org/10.1086/340549}
	
	\bibitem{mandal2023cosmic} S. Mandal, A. Singh, R. Chaubey, Cosmic evolution of holographic dark energy in $f(Q, T)$ gravity, International Journal of Geometric Methods in Modern Physics, \textbf{20}, 2350084 (2023) \url{https://doi.org/10.1142/S0219887823500846}
	\bibitem{scolnic2018complete} D. M. Scolnic, D. O. Jones, A. Rest, Y. C. Pan, R. Chornock et al., The complete light-curve sample of spectroscopically confirmed SNe Ia from Pan-STARRS1 and cosmological constraints from the combined Pantheon sample, The Astrophysical Journal, \textbf{859}, 101 (2018) \url{https://doi.org/10.3847/1538-4357/aab9bb}
	\bibitem{riess1999bvri} A. G. Riess, R. P. Kirshner, B. P. Schmidt, S. Jha, P. Challis et al., BVRI light curves for 22 type Ia supernovae, The Astronomical Journal, \textbf{117}, 707 (1999) \url{https://doi.org/10.1086/300738}
	\bibitem{hicken2009improved} M. Hicken, W. M. Wood-Vasey, S. Blondin, P. Challis, S. Jha et al., Improved dark energy constraints from ~100 new CfA supernova type Ia light curves, The Astrophysical Journal, \textbf{700}, 1097 (2009) \url{https://doi.org/10.1088/0004-637X/700/2/1097}
	\bibitem{sako2018data} M. Sako, B. Bassett, A. C. Becker, P. J. Brown, H. Campbell et al., The data release of the Sloan Digital Sky Survey-II supernova survey, Publications of the Astronomical Society of the Pacific, \textbf{130}, 064002 (2018) \url{https://doi.org/10.1088/1538-3873/aab4e0}
	\bibitem{guy2010supernova} J. Guy, M. Sullivan, A. Conley, N. Regnault, P. Astier, et al., The Supernova Legacy Survey 3-year sample: Type Ia supernovae photometric distances and cosmological constraints, Astronomy $\&$ Astrophysics, \textbf{523},  (2010) \url{https://doi.org/10.1051/0004-6361/201014468}
	\bibitem{contreras2010carnegie} C. Contreras, M. Hamuy, M. M. Phillips, G. Folatelli, N. B. Suntzeff, et al., The Carnegie Supernova Project: first photometry data release of low-redshift type Ia supernovae, The Astronomical Journal, \textbf{139}, 519 (2010) \url{https://doi.org/10.1088/0004-6256/139/2/519}
	\bibitem{odintsov2018cosmological} S. D. Odintsov, V. Oikonomou, A. Timoshkin, E. N. Saridakis, R. Myrzakulov, Cosmological fluids with logarithmic equation of state, Annals of Physics, \textbf{398}, 238--253 (2018) \url{https://doi.org/10.1016/j.aop.2018.09.015}
	\bibitem{ellis2012relativistic} G. F. R. Ellis, R. Maartens, M. A. H. MacCallum, Relativistic cosmology, Cambridge University Press, (2012) \url{http://dx.doi.org/10.1017/CBO9781139014403}	
	
	\bibitem{asvesta2022observational} K. Asvesta, L. Kazantzidis, L. Perivolaropoulos, C. G. Tsagas, Observational constraints on the deceleration parameter in a tilted universe, Monthly Notices of the Royal Astronomical Society, \textbf{513}, 2394--2406 (2022) \url{https://doi.org/10.1093/mnras/stac922}	
	\bibitem{Zhao2006} W. Zhao, Y. Zhang, Quintom models with an equation of state crossing-1, Physical Review D, \textbf{73}, 123509 (2006) \url{https://doi.org/10.1103/PhysRevD.73.123509}
	\bibitem{visser1997energy} M. Visser, Energy conditions in the epoch of galaxy formation, Science, \textbf{276}, 88--90 (1997) \url{https://doi.org/10.1126/science.276.5309.88}
	\bibitem{lalke2024cosmic} A. R. Lalke, G. P. Singh, A. Singh, Cosmic dynamics with late-time constraints on the parametric deceleration parameter model, European Physical Journal Plus, \textbf{139}, 288 (2024) \url{https://doi.org/10.1140/epjp/s13360-024-05091-5}
	\bibitem{singh2022lagrangian} A. Singh, R. Raushan, R. Chaubey, S. Mandal, K. C. Mishra, Lagrangian formulation and implications of barotropic fluid cosmologies, International Journal of Geometric Methods in Modern Physics, \textbf{19}, 2250107 (2022) \url{https://doi.org/10.1142/S0219887822501079}
	\bibitem{mishra2025cosmological} S. S. Mishra, P. K. Sahoo, Cosmological aspects in the constrained $f(T, T)$ theory using Raychaudhuri equations, Physics of the Dark Universe, \textbf{48}, 101887 (2025) \url{https://doi.org/10.1016/j.dark.2025.101887}
	\bibitem{Sahni2003} V. Sahni, T. D. Saini, A. A. Starobinsky, U. Alam, Statefinder-a new geometrical diagnostic of dark energy, Journal of Experimental and theoretical Physics Letters, \textbf{77}, 201–206 (2003) \url{https://doi.org/10.1134/1.1574831}
	\bibitem{fei2013statefinder} Y. Fei, Z. Jing-Fei, Statefinder diagnosis for the extended holographic Ricci dark energy model without and with interaction, Communications in Theoretical Physics, \textbf{59}, 243--248 (2013) \url{https://doi.org/10.1088/0253-6102/59/2/17}
	\bibitem{Akbar2006er} M. Akbar, R. G. Cai, Friedmann equations of FRW universe in scalar-tensor gravity, $f(R)$ gravity and first law of thermodynamics, Phys. Lett. B, \textbf{635}, 7--10 (2006) \url{https://doi.org/10.1016/j.physletb.2006.02.035}
	\bibitem{bamba2012reconstruction} K. Bamba, R. Myrzakulov, S. Nojiri, S. D. Odintsov, Reconstruction of $f(T)$ gravity: Rip cosmology, finite-time future singularities,<? format?> and thermodynamics, Physical Review D, \textbf{85}, 104036 (2012) \url{https://doi.org/10.1103/PhysRevD.85.104036}
	\bibitem{tong2009cosmic} M. L. Tong, Y. Zhang, Cosmic age, statefinder and Om diagnostics in the decaying vacuum cosmology, Physical Review D, \textbf{80}, 023503 (2009) \url{https://doi.org/10.1103/PhysRevD.80.023503}
	
	
	 	 
\end{thebibliography}
\end{document}